\documentclass[iop]{emulateapj}
\usepackage{amsmath}
\RequirePackage{lineno}
\usepackage{graphicx}

\usepackage{enumerate}
\usepackage{cases}

\newcommand{\swift}{\emph{Swift}}
\newcommand{\fermi}{\emph{Fermi}}

\newcommand{\e}[1]{10$^{#1}$}
\newcommand{\ee}[1]{$\times$10$^{#1}$}

\newcommand{\rate}{~cts\,s$^{-1}$}
\newcommand{\flux}{~erg\,cm$^{-2}$\,s$^{-1}$}
\newcommand{\fluence}{~erg\,cm$^{-2}$}

\shorttitle{X-ray flares with Swift and Fermi}
\shortauthors{Troja et al.} 

\begin{document}
\title{\textit{Swift} and \textit{Fermi} observations of X-ray flares: the case of Late Internal shock}


\author{E. Troja\altaffilmark{1,2}, L. Piro\altaffilmark{3}, 
V.~Vasileiou\altaffilmark{4}, N.~Omodei\altaffilmark{5}, 
J.~M.~Burgess\altaffilmark{6}, S.~Cutini\altaffilmark{7},
V.~Connaughton\altaffilmark{6}, J.~E.~McEnery\altaffilmark{8}}
\altaffiltext{1}{Center for Research and Exploration in Space Science and
Technology, NASA Goddard Space Flight Center, Greenbelt, MD 20771, USA}
\altaffiltext{2}{Department of Astronomy, University of Maryland, 
College Park, MD 20742.}
\altaffiltext{3}{INAF - IAPS, Via Fosso del Cavaliere 100, 00133 Rome, Italy}
\altaffiltext{4}{Laboratoire Univers et Particules de Montpellier, 
Universite Montpellier 2, and CNRS/IN2P3, Montpellier, France}
\altaffiltext{5}{W. W. Hansen Experimental Physics Laboratory, 
Kavli Institute for Particle Astrophysics and Cosmology, 
Department of Physics and SLAC National Accelerator Laboratory, 
Stanford University, Stanford, CA 94305, USA}
\altaffiltext{6}{University of Alabama in Huntsville, 
NSSTC, 320 Sparkman Drive, Huntsville, AL 35805, USA}
\altaffiltext{7}{ASI Science Data Center, via Galileo Galilei, 
00044 Frascati, Italy}
\altaffiltext{8}{Astrophysics Science Division, 
NASA Goddard Space Flight Center, Greenbelt, MD 20771}

\begin{abstract}
Simultaneous {\it Swift} and {\it Fermi} observations of gamma-ray bursts 
(GRBs) offer a unique broadband view of their afterglow emission, 
spanning more than ten decades in energy.
We present the sample of X-ray flares observed by both {\it Swift} and 
{\it Fermi} during the first three years of {\it Fermi} operations. 
While bright in the X-ray band, X-ray flares are often 
undetected at lower (optical), and higher (MeV to GeV) energies.  
We show that this disfavors synchrotron self-Compton processes as 
origin of the observed X-ray emission.
We compare the broadband properties of X-ray flares with the standard
late internal shock model, and find that, in this scenario, 
X-ray flares can be produced by a late-time relativistic ($\Gamma$$>$50) 
outflow at radii $R$$\sim$10$^{13}-10^{14}$\,cm. This conclusion holds
only if the variability timescale is significantly shorter than the observed
flare duration, and implies that X-ray flares can directly probe the
activity of the GRB central engine.

\end{abstract}



\maketitle

\section{Introduction}

Gamma-ray bursts (GRBs) are extremely energetic events, 
releasing a significant amount of energy ($\gtrsim$\e{51}\,erg) 
over a short timescale (seconds to minutes) in the form of highly relativistic jets. 
The nature of the astrophysical source powering such energetic outflows 
-- the so-called central engine -- is still unsettled. 
The GRB central engine is in fact hidden from direct
probing with photons, and it can be potentially accessed only by means of 
neutrino or gravitational wave measurements 
\citep{fryer03,suwa09}. 
Nevertheless, inferences on its nature can be made through the study of GRBs and the
 high-energy properties of their afterglows
\citep[e.g.][]{troja07,lyons10,lu14}.
In this respect, X-ray flares represent one of the most promising diagnostic tools
for tracing the time history of the central engine activity \citep{maxham09}.

X-ray flares, glimpsed at by {\it Beppo-SAX} in a few events \citep{piro98,piro05,galli06}, 
were revealed only by \swift~\citep{swift} as a common feature of GRB afterglows 
\citep{burrows05}.
They appear as sudden brightening episodes, often 
characterized by rapid rise and decay times ($\Delta t/ t \approx$0.2; 
\citealt{chinca07}), 
during which the X-ray flux increases by a factor of several
to hundreds \citep{falcone06}.
X-ray flares are commonly observed in long GRBs ($\sim$30\%; \citealt{chinca07})
and, to a lesser extent, in short GRBs \citep{barthelmy05,laparola06}.
They are observed in all phases of the afterglow \citep{obrien06,nousek06}, 
typically $\sim$100 s to $\sim$1000 s after the burst, and sometimes 
extending to over a day. 
GRB afterglows usually exhibit one or two flares, but multiple flares are also sporadically 
observed \citep[e.g.][]{perri07,100728a}.

The discovery of X-ray flares triggered a number of theoretical studies,
relating the X-ray emission to the external forward/reverse shock \citep{galli06,koba07,guetta07,pana08,mesler12},
late internal shock \citep{fan05,zhang06}, jet propagation instabilities
\citep{lazzati11}, or a delayed magnetic dissipation \citep{giannios06}.
The most important difference between these models is that some of them 
require a long duration and/or a re-activation of the central engine,
e.g. through erratic accretion episodes \citep{king05} or accretion disc instabilities
\citep{proga06},
while some others do not \citep[e.g.][]{piro05,giannios06,belo11}. 
Distinguishing between these scenarios bears important implications 
for the physics of the GRB central engine, and for the 
energy extraction mechanism that powers the observed high-energy emission. 

\swift~observations provided fundamental clues on the nature of X-ray flares.
In particular, the observed rapid variability challenges most (but not all) 
of the emission mechanisms related to the external shock emission \citep{lazzati07}. 
The X-ray temporal and spectral properties of flares \citep{chinca07,falcone07}
 hint at a direct link with the prompt gamma-ray emission, 
 and suggest a common origin of the two phenomena,
but they do not definitely break the degeneracy between the different scenarios. 
However, in a wide range of models the X-ray flare is accompanied by a 
higher energy counterpart peaking in the MeV/GeV energy range
\citep{wang06,galli07,fan08,yu09}. Different models predict substantially different properties 
of the high-energy ($>$MeV) emission, 
which could therefore provide a discriminating observational feature. 

A critical factor in shaping the high-energy emission associated with X-ray flares
is the location of the emitting region: internal, $R$$\approx$\e{13}-\e{15}\,cm, or external, $R$$\gtrsim$\e{17}\,cm. 
In a first group of models, X-ray flares are produced by means similar to those which produce the prompt emission (i.e. 
internal shocks, or some other dissipation process within the ultrarelativistic outflow), 
but at later times and at lower energies. In this scenario the observed X-ray emission is 
generally attributed to synchrotron radiation, whereas high-energy flares are produced by 
synchrotron self-Compton (SSC) scattering of the X-ray photons, or
through late inverse Compton (IC) scattering by electrons accelerated at the external radius.
In the SSC case, the X-ray and high-energy flares are generated in the same region and 
by the same electron population. They are therefore expected to 
be temporally correlated. 
As the peak of the flare spectrum is in the EUV/X-ray range, 
the peak energy of the SSC component produced by internal shocks is expected to lie 
around $\approx$10-100~MeV \citep{gg03,wang06}.
In the case of the external inverse Compton (EIC) process, 
the X-ray flare photons need time to reach and scatter with the afterglow electrons. 
During that time the beam spreads out, and this causes a delayed and longer lasting high-energy flare.
As external shocks are characterized by much larger radii, 
the spectral peak of the EIC component is expected to fall in the GeV band \citep{wang06,he11}.  

An alternative set of models suggest that X-ray flares are produced
by the interaction of the relativistic outflow with the external medium.
In this scenario the X-ray photons are up-scattered through first-order or second-order IC processes at the deceleration radius, 
thus producing a bright flare in the GeV-to-TeV band. The low- and high-energy flares 
show similar temporal profiles
and no significant delay \citep{galli07}.

The unprecedented broadband coverage (from a few eV to tens of GeV) of simultaneous \swift~and \fermi~observations
offers for the first time the opportunity to test these predictions, and 
to fully exploit the observations of X-ray flares. 
We systematically searched the \fermi~data for high-energy emission associated with the 
X-ray flares detected by \swift. 
GeV emission was detected during the X-ray flares of GRB~100728A, 
as reported in \citet{100728a}.
In this paper we report the  results of our search on the whole sample of X-ray flares, 
and compare the broadband spectra of flares with the predictions of the 
late internal shock model. 
The paper is organized as follows: our selection criteria are listed in \S~\ref{sample};
data reduction and analysis are described in \S~\ref{swift}-\ref{lat};
the theoretical model is derived in \S~\ref{sec:model}; the results of the 
search for HE flares and of the broadband spectral fits are discussed in 
\S~\ref{sec:results}. Throughout the paper times are measured from the 
{\it Swift} trigger time. Unless otherwise stated, uncertainties are quoted at the 90\%
confidence level for one parameter of interest. 
\newpage
\section{Data Analysis}
\subsection{Sample Selection}\label{sample}

We visually inspected all the \swift~afterglow light curves observed during
the first three years of \fermi\ operations, 
between  2008 August 01 and 2011 August 01,  
and selected our sample according to the following criteria:

\begin{enumerate}[I.]
\item The X-ray light curve shows a significant rebrightening, 
the peak flux being at least a factor of $\sim$3 higher than the underlying continuum.

\item The flare peaks at early times (t$_{\rm pk}$$<$1000\,s).
Flares peaking at later times are less frequent and typically fainter \citep{curran08}, 
and {\it Fermi} observations for such low-flux flares are not constraining.

\item During the flare time interval the GRB location is within the field of view (FoV)
of the Large Area Telescope (LAT; \citealt{atwood09}), 
i.~e. the GRB angle to the LAT boresight is $\theta_{\rm LAT}<65^{\circ}$, 
and it is not occulted by the Earth, i.~e. the zenith angle is $\theta_{\rm z}<95^{\circ}$. 

\item The flare follows a bright prompt emission. This last constraint was
introduced to avoid cases in which \swift~triggered on a weak 
precursor \citep[e.g.][]{page07}, and the rebrightening observed in the X-ray band
corresponds to the main prompt emission rather than a typical X-ray flare.

\end{enumerate}

During the 3 yr period here considered \swift\ detected 264 GRBs. 
Among them 77 bursts
do not have early time (t$<$1000\,s) X-ray observations, either because an observing constraint prevented a prompt slew or because the burst was detected in ground analysis, and therefore are not relevant for this work. 
In the remaining sample, 55 bursts ($\sim$30\%) show early time X-ray flares as defined in (I) and (II). 
We note that only one (GRB~100117A) is classified as a short GRB. 
For a sizable numbers of GRBs (14 out of 55) there are good LAT observations, as defined in (III), during the flare time interval. 
By applying our criterion (IV) we exclude from this sample GRB~090621A.
The subsample of events with measured redshift,
either from afterglow spectroscopy or photometry, comprises five bursts:
GRB~080906, GRB~080928, GRB~081203A, GRB~090516, and GRB~110731A.

\begin{table*}[!thb]
\label{tab:obslog}
\begin{center}
\caption{Properties of the sample of early X-ray flares}
\begin{tabular}{lccccccccccc}
\hline
{GRB\ \ \ \ \ \ \ \ \ \ \ \ \ \ \ \ \ \ \ \ \ \ \ \ \ \ \ \ } &
$T_{90}$  & $S_{\gamma}$  &
$t_{\rm pk}$ &  $\Delta t$ &
Model & $\alpha$ & $\Gamma$/$\beta$ & $E_{pk}$ &
$F_{\rm X}$ & $\chi^2$/d.o.f.\\
(1) & (2) & (3) &
(4) & (5) &
(6) & (7) & (8) &
(9) & (10) & (11) \\
\hline
090407\dotfill  & 310$\pm$70  	& 11$\pm$2    
                & 138 		& 34  
		& PL 		& -- 	& 1.84$\pm$0.06 	& -- 
		& 4.2$\pm$0.2 &  129/125 \\		
090831C\dotfill 	& 3.3$\pm$1.0 	& 1.5$\pm$0.3 
 			& 186 		& 78  
			& PL 		& -- 	& 1.9$^{+0.7}_{-0.5}$ 	& -- 
			& 0.35$\pm$0.17 & 50/58$^a$ \\
091221\dotfill  	& 69$\pm$6    	& 57$\pm$2    
			& 108 		& 38  
			& PL 		& -- & 2.35$\pm$0.14 	& -- 
			& 1.9$\pm$0.2	& 68/66 \\

100212A\dotfill 	& 136$\pm$14  	& 9.1$\pm$1.2 
			&  80 		&  8  
			& Band 		&  1.24$\pm$0.12 & $>$1.8 	& 27$^{+20}_{-12}$    
			& 9.8$\pm$0.2 	&  91/88  \\
                 	&         	&             
			& 118 		& 50  
			& Band 		& 0.95$\pm$0.15   & 2.5$\pm$0.2  & 3.5$\pm$0.4
			& 11.4$\pm$0.6  & 194/210 \\
		 	&         	&             
			& 226 		& 20  
			& Band 		& 0.6$\pm$0.4	& 3.4$^{+0.4}_{-1.3}$ & 1.55$\pm$0.11         
			& 2.9$\pm$0.3   &  47/60  \\
                 	&         	&             
			& 250 		& 25  
			&  Band 	& 1.2$\pm$0.3  	& $>$2.9 & 1.12$\pm$0.14 
			& 2.5$\pm$0.3 	&  69/65  \\
		 	&         	&             
			& 353 		& 27  
			&  PL 		& -- 	& 2.54$\pm$0.14 & -- 
			& 1.85$\pm$0.25 &   53/47  \\
		 	&         	&             
			& 423 		& 27  
			&  PL 		& --  	& 2.9$\pm$0.2   & -- 
			& 0.98$\pm$0.03 & 182/189$^a$ \\
		 	&         	&             
			& 665 		& 77  
			&  PL 		& -- 	& 2.9$\pm$0.5   & -- 
			& 0.35$\pm$0.15 &  91/104$^a$ \\
100725B\dotfill 	& 200$\pm$30  	& 68$\pm$2    
			& 135 		& 46  
			& Band & 0.74$\pm$0.15 & 1.86$\pm$0.05 & 22$^{+5}_{-4}$    
			& 25.1$\pm$1.0 	&  187/198 \\
			&     		&             
		  	& 162 		& 62  
			& Band &  1.10$^{+0.12}_{-0.20}$ &   2.9$^{+0.3}_{-0.2}$  & 6.8$^{+0.7}_{-0.8}$    
			& 25$\pm$2 	& 188/205    \\
          		&            	&             		
	  		& 215 		& 29  
			& Band 		&  1.50$^{+0.19}_{-0.4}$ & 2.76$^{+0.3}_{-0.16}$  
			& 3.0$\pm$0.5    
			& 26.0$\pm$0.9 	&   141/164 \\
          		&            	&             
			& 270 		& 20  
			& Band & 1.5$^{+0.4}_{-1.7}$ & $>$3 & 0.8$^{+0.6}_{-0.5}$
			& 7$\pm$2 	&    118/117  \\
110102A\dotfill 	& 264$\pm$8	&  165$\pm$3
			& 211	&  50
			& Band	& 0.65$^{+0.3}_{-0.4}$	&  1.46$\pm$0.03 
			& 10$^{+5}_{-3}$
			& 29.5$\pm$1.0	& 208/191 \\
 			& 	& 
			& 263	& 50 
			& PL	& -- 	&  1.55$\pm$0.03
			& -- 
			& 5.8$\pm$0.2	& 329/281 \\			  
110414A\dotfill & 152$\pm$73 & 35$\pm$3    &  354   &  230   & PL & -- &  2.02$^{+0.19}_{-0.16}$ & -- & 0.15$\pm$0.03 &
177/224$^a$ & \\ 

\hline
\multicolumn{11}{c}{Bursts with known redshift} \\
\hline
080906 ($z$=2.0)\dotfill  & 150$\pm$20 & 35$\pm$2 & 178 &  59 &PL & -- & 1.97$\pm$0.08 & -- & 1.5$\pm$0.2 & 73/90 \\
                           &            &          & 613 & 281 &PL &-- & 1.83$\pm$0.15 & -- & 0.19$\pm$0.05 &  332/354$^a$  \\
			   
080928 ($z$=1.69)\dotfill 	& 280$\pm$30 & 25$\pm$2 
					& 354 & 34 
					& Band & 0.65$^{+0.5}_{-0.4}$ & 2.0$\pm$0.2 & 2.7$^{+1.0}_{-0.9}$
					& 1.80$\pm$0.5 & 127/119 \\					
081203A($z$=2.1)\dotfill  	& 220$\pm$90 	& 78$\pm$3 
					&  89 &  29 
					& Band+PL & 0.3$^{+0.3}_{-0.5}$ & 3.2$\pm$0.5 & 1.1$^{+0.11}_{-0.05}$ 
					& 3.8$\pm$0.2 & 225/206  \\
					
090516 ($z$=4.1)\dotfill 	& 210$\pm$65 & 90$\pm$6 
					& 275 &  30 
					&Band & 1.20$^{+0.18}_{-0.15}$ & 2.5$^{+0.26}_{-0.17}$ 
					& 2.4$\pm$0.7 & 10.3$\pm$1.0 & 141/139 \\

110731A  ($z$=2.83)\dotfill 	& 39$\pm$13	 &  60$\pm$1	 
					& 70	 &  27	 
					& PL	 &  --	 &  1.96$\pm$0.05 & --	 
					& 3.9$\pm$0.2	 &  109/102 \\

\hline
\end{tabular}

\begin{minipage}[h]{0.87\linewidth}
\vspace{0.05in}
$^a$~Low counts spectra were rebinned in order to have at least 1 count in each energy channel. The best fit model was found by minimizing 
the Cash statistic.

\vspace{0.05in}
{\sc Notes :} Col. (1): GRB name;
Col. (2): T$_{90}$ duration (in s) in the 15-350~keV energy band;
Col.~(3): burst fluence (in units of \e{-7} \fluence) in the 15-150~keV energy band ;
Col.~(4): peak time (in s) of the X-ray flare;
Col. (5): temporal width (in s) of the X-ray flare;
Col. (6): best fit model: a power law (PL), or a Band function (Band);
Col. (7): low-energy photon index of the Band function;
Col. (8): photon index of the PL fit/ high-energy photon index of the Band function;
Col. (9): peak energy (keV);
Col. (10): unabsorbed X-ray flux (in units of \e{-9}\,erg\,cm$^{-2}$\,s$^{-1}$) in the 0.3-10 keV energy band;
Col. (11): chi-squared over d.o.f.
\end{minipage}
\end{center}
\end{table*}

\subsection{\textit{Swift} data}\label{swift}

The \swift~data were retrieved from the public archive\footnote{
\href{http://heasarc.gsfc.nasa.gov/docs/swift/archive/}
{http://heasarc.gsfc.nasa.gov/docs/swift/archive/}}
and processed with the standard \swift~analysis software (v3.9)
included in NASA's HEASARC software (HEASOFT, ver.~6.12)
and the relevant calibration files.

Burst Alert Telescope (BAT; \citealt{bat05}) mask-weighted light curves and spectra were
extracted in the nominal 15-150 keV energy range 
following the standard procedure.
The automatic background subtraction performed by the
\swift~software
is correct only if there are no other bright hard X-ray sources
in the BAT FoV. When this condition is not satisfied, a systematic
contamination of the source fluxes arises.
The residual background contribution is negligible during the main prompt emission,
 but in the case of X-ray flares the signal detected in the
BAT energy range, if any, is usually very weak.
The presence of nearby sources may significantly
affect the count rate derived through the mask-weighting technique.
In order to properly remove this residual background term, we
used the tool \textsc{batclean} inputting the position of the known
hard X-ray sources in the FoV to create a background map
for each spectral channel.
The background map and the coded mask pattern were used
to reconstruct the sky images (tool \textsc{batfftimage}).
The correct source count rate in each energy channel
were derived from the sky images with the tool \textsc{batcelldetect}.

X-Ray Telescope (XRT; \citealt{xrt05}) light curves and spectra were extracted in the nominal 0.3-10 keV energy
range by applying standard screening criteria. All the XRT data products
presented here are background subtracted and corrected for
point-spread function (PSF) losses, vignetting effects and exposure variations.
We refer the reader to \citet{evans07,evans09}
for further details on the XRT data reduction.
The X-ray light curves were fit with one or more power-law segments
describing the smooth afterglow decay \citep{nousek06}, superposed with a power-law rise/
exponential decay profile for any flares.
We defined the flare width $\Delta t$ as the time interval between the $1/e$ intensity points \citep[e.g.][]{chinca07}.
The best fit model was used to determine the start and stop times of the flares, defined as the times that the flare profile intersects the residual curve (continuum + other flares). Spectroscopy and the search for
high-energy emission (\S~\ref{lat}) were perfomed in this time interval.

Spectral fits were performed with XSPEC~v.~12.7.1 \citep{arnaud96}.
Unless otherwise stated, the X-ray spectra were binned to
$>$20~counts/bin and $\chi^2$ statistics were used. If the
X-ray flare was also detected by BAT, the BAT and XRT spectra were
jointly fit by holding the normalization between the two
instruments fixed to unity. Two absorption components were
included: one fixed at the Galactic value
\citep{kalberla05}, and the other, representing the absorption local
to the burst, was constrained from the late-time afterglow
spectroscopy \citep{evans09}. The flares spectra can be well
described by a simple power law, or a smoothly broken power law 
(Band function; \citealt{band93}). In only one case (GRB~081203A)
is the resulting fit poor ($\chi^2$/d.o.f.=394/210 for a power-law model, 
and $\chi^2$/d.o.f.=323/208 for a Band model) as it underestimates by a 
factor of $\sim$4 the observed flux in the BAT energy range. 
The addition of a power law with
photon index $\Gamma$=1.5$\pm$0.2 yields a significant improvement
($\chi^2$/d.o.f.=225/206). 
The temporal and spectral properties of
the sample of X-ray flares are summarized in Table~1.


\begin{figure*}[!tb]
\centering
\includegraphics[angle=270,scale=0.46]{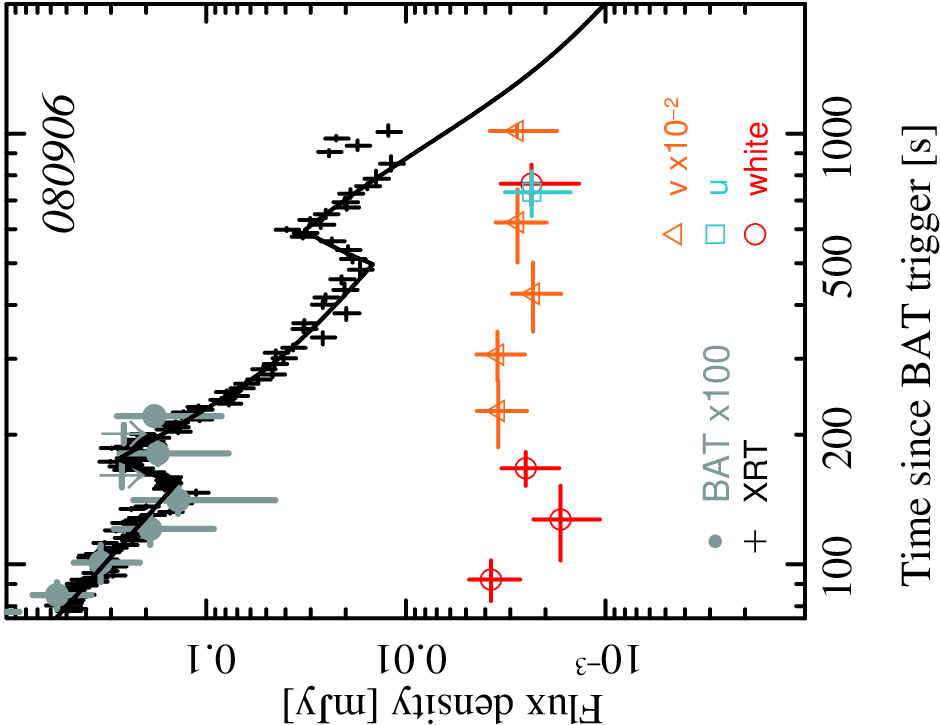}
\hspace{0.2cm}
\includegraphics[angle=270,scale=0.46]{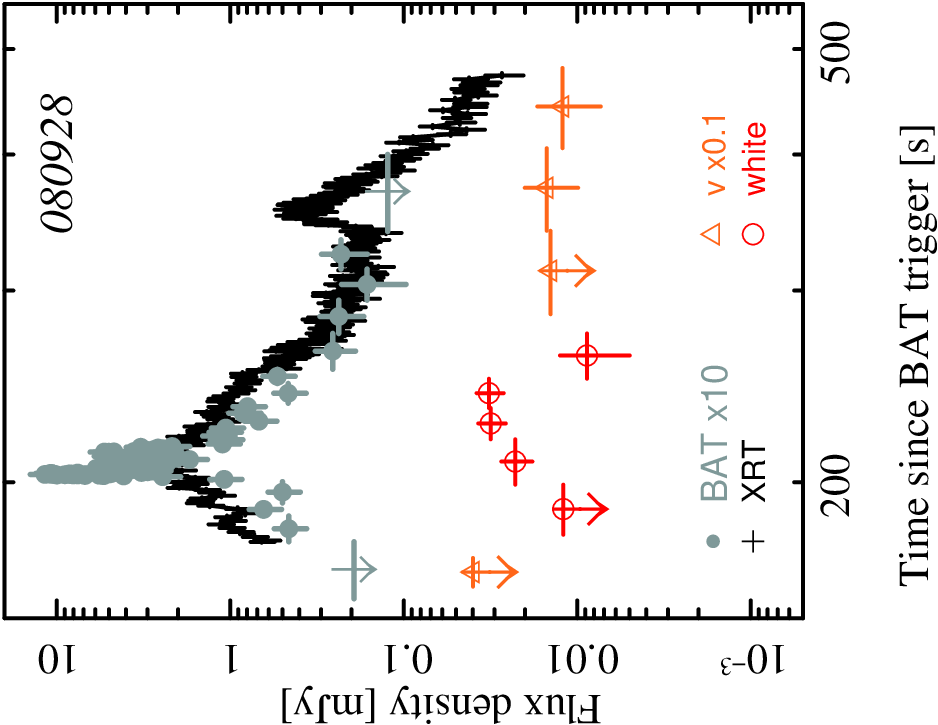}
\hspace{0.2cm}
\includegraphics[angle=270,scale=0.46]{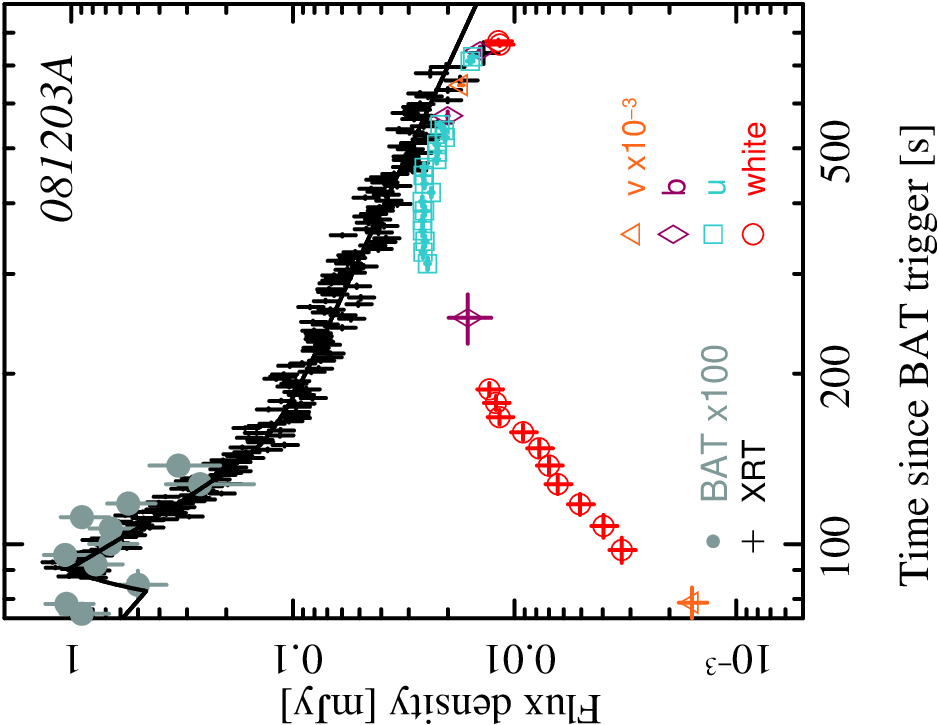}\\
\vspace{0.4cm}
\includegraphics[angle=270,scale=0.46]{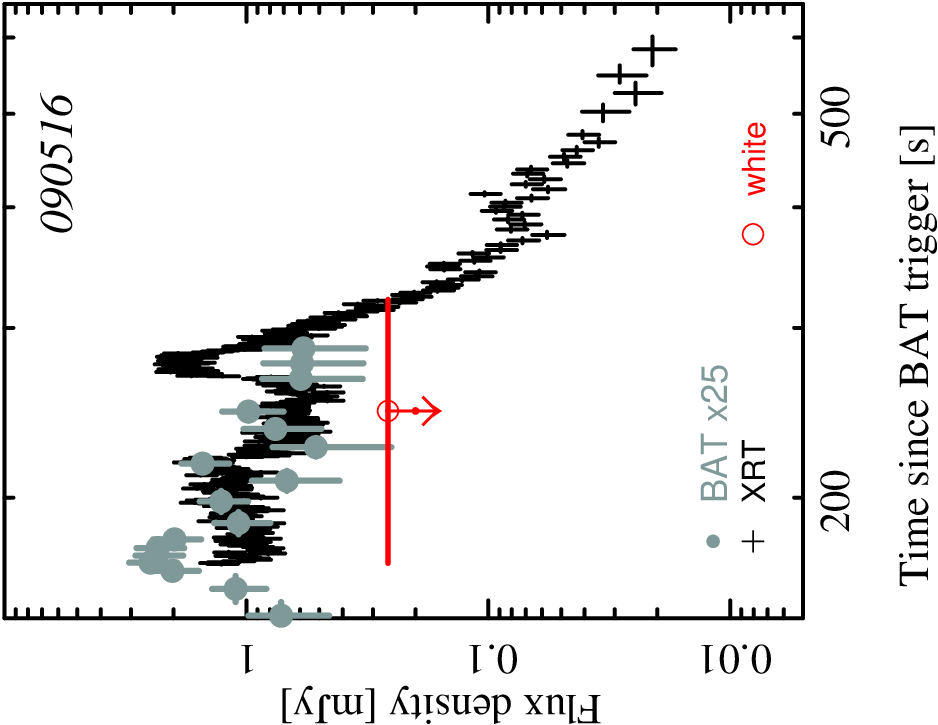}
\hspace{0.2cm}
\includegraphics[angle=270,scale=0.46]{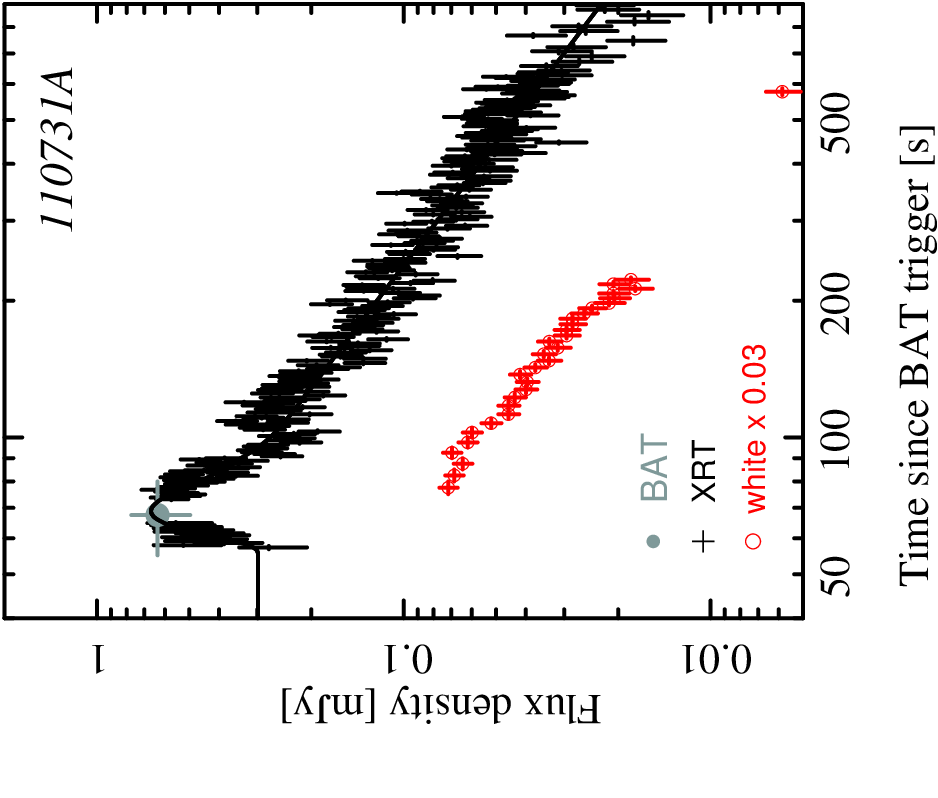}\\

\caption{\swift~multi-wavelength light curves for the Gold Sample of X-ray flares.
X-ray fluxes are corrected for Galactic and intrinsic absorption.
UVOT fluxes have been renormalized to the V-band,
and corrected for extinction.}
\label{fig:swiftlc}
\end{figure*}


UV/Optical Telescope (UVOT; \citealt{uvot05}) photometric measurements were performed on a circular source extraction region
with a radius of 5\arcsec. In case of faint ($\lesssim$0.5\rate) sources a 3\arcsec
~radius aperture was used in order to optimize the signal-to-noise ratio and an
aperture correction factor was applied for consistency with the UVOT
calibration \citep{breeveld10,breeveld11}. 
The data were corrected for Galactic extinction \citep{schlegel98}
and, when possible, for intrinsic host extinction.

The \swift~multi-wavelength light curves for the subsample of bursts with known redshift,
which will be used for our broadband spectral modeling, are shown in Fig.~\ref{fig:swiftlc}.

\subsection{\textit{Fermi}/GBM data}\label{sec:gbm}

We examined the daily Gamma-Ray Burst Monitor (GBM) CSPEC data\footnote{Available at the \textit{Fermi}'s Science 
Support Center: ftp://legacy.gsfc.nasa.gov/fermi/data/gbm/daily/} searching for
higher energy ($>$300\,keV)
counterparts of the X-ray flares. 
We considered the two GBM  bismuth germanate (BGO) detectors
and for the time of each flare selected the detector with the best view of the burst, i.e. 
with the smaller angle between the GRB's direction and the detector. 
For each burst we examined the light curve in the 300\,keV--10\,MeV energy range.
The expected number of background events (N$_\mathrm{exp}$)  
was estimated by fitting a 2nd or 3rd order polynomial to the pre-flare and post-flare data,
and by interpolating the fit during the X-ray flare time interval.
The source significance was calculated as  
\mbox{$\mathcal{S}=(\mathrm{N_{det}}-\mathrm{N_{exp}})/\sqrt{\mathrm{N_{exp}}}$},
where N$_\mathrm{det}$ is the number of events detected during the flare time interval.

Only in the case of GRB~110102A, a significant detection (6.9$\,\sigma$) was found 
during the first flare.  
The signal is detected only up to 1\,MeV, and 
is compatible with the extrapolation of the flare keV emission to higher energies.
No significant increase of the event rate was visible in coincidence with any
other X-ray flares. 

By using a Bayesian approach with a flat prior $\pi(S)=1$ for $S>0$ and 0 otherwise on the signal S \citep{nakamura10}, 
we calculated an upper limit of confidence level CL on the number of signal events $S_\mathrm{UL}$ by numerically solving
the following equation:
\begin{align}
CL&=\frac{\int^{S_\mathrm{UL}}_{-\infty} P_{G}(\mathrm{N_{det}},\mathrm{N_{exp}}+S)\pi(S)dS}{\int^{\infty}_{-\infty} P_{G}(\mathrm{N_{det}},\mathrm{N_{exp}}+S)\pi(S)dS} \nonumber \\
&= \frac{\int^{S_\mathrm{UL}}_{0} P_{G}(\mathrm{N_{det}},\mathrm{N_{exp}}+S)dS}{\int^{\infty}_{0} P_{G}(\mathrm{N_{det}},\mathrm{N_{exp}}+S)dS}
\end{align}
where $P_G(n,s)$ is the Poisson probability of detecting $s$ events when expecting $n$.
We assumed no systematic uncertainties on the estimated number of background events N$_\mathrm{exp}$. 
For each flare we generated the response matrix of the selected BGO detector 
by using \textsc{GBM\_RSP\_Gen} v.~1.8,
and derived the detector effective area as a function of energy 
by assuming no energy-dispersion effects. 
A spectrally-weighted effective area, $A_\mathrm{eff}$, was calculated for a 
photon index $\Gamma$=2.0, and used to convert the upper limits into flux units. 
The derived values are listed in Table~\ref{gbmul}. 
In the case of a spacecraft autonomous repointing (GRB~110731A, and GRB~110414A), 
the rapid change of the spacecraft's orientation did not allow us to properly estimate the background level,
and upper limits are not reported.

\begin{table}[!t]
\begin{center}
 \caption{GBM Upper Limits}
\begin{tabular}{lcccc}
\hline 
GRB\ \ \ \ \ \ \ \ \ \ \ \ \ \  & \ \ $t_{\rm i}$ \ \ & \ \ $t_{\rm f}$\ \  
& \ \ \ $F_{\rm 300 keV -1\,MeV}$\ \ \ & \ \ \ $F_{\rm 1-10\,MeV}$ \ \ \  \\
(1) & (2) & (3) & (4) & (5) \\
 \hline
080906\dotfill & 160   &   240 & 2.4 & 2.3 \\
       & 520   &   800 & 1.4 & 1.1 \\
080928\dotfill & 340   &   400 & 5.2 & 4.4 \\
081203A\dotfill & 98   &   154 & 3.8 & 4.5 \\
090407\dotfill &120   &   170 &4.3 & 4.9\\
090516\dotfill & 260   &   315 & 6.0 & 4.9 \\
090831C\dotfill &160   &   260 & 3.4 & 3.7 \\
091221\dotfill & 95   &   160 & 4.0 & 4.2  \\
100725B\dotfill & 115   &   150 & 6.9 & 7.5\\
       & 150   &   200 & 5.7& 6.3\\
       & 200   &   255 & 5.4& 7.0\\
       & 255   &   315 & 5.1& 5.7\\
100212A\dotfill & 76   &   85 & 10.3 & 11.3\\
       & 95   &   160& 3.9 &  4.2\\
       & 223   &   243 & 7.0 & 7.6\\
       & 249   &   275 & 6.1& 6.7\\
       & 344   &   383 & 5.1 & 5.5\\
       & 413   &   445 & 5.7 & 6.1\\       
       & 628   &   750 &3.1 & 3.3\\
110102A\dotfill &249   &   393 &    --\footnote{There was significant signal detection in this interval} & 4.7\\
       &249   &   393 & 2.7 & 2.8\\
\hline
\label{gbmul}
\end{tabular}
\end{center}
\vspace{-0.4cm}
{\sc Notes :} Col. (1): GRB name;
Col.~(2): start time of the search, in units of s;
Col.~(3): stop time of the search, in units of s;
Col. (4): 95\% upper limit in the 300~keV-1~MeV band. Units are \e{-2}\,ph\,cm$^{-2}$\,s$^{-1}$.
Col. (5): 95\% upper limit in the 1~MeV-10~MeV band. Units are \e{-2}\,ph\,cm$^{-2}$\,s$^{-1}$.
\end{table}

\subsection{\textit{Fermi}/LAT data}\label{lat}

The LAT data were searched for emission related to the X-ray flares observed by \swift.  
We searched for emission coincident in time with each X-ray flare.  
In the case of a GRB afterglow with multiple flares, the search was also performed by stacking the data of the whole flaring activity, 
that is one search for the seven flares of GRB100212A, one for the four flares of GRB 100725B, one for the eight flares in GRB~100728A and one for the 2 flares of GRB 080906. 
According to some models, the high-energy emission could be delayed and longer-lasting than the lower energy flare. We therefore searched the LAT data for emission 
over longer time scales, 
performing our search over a period from the start time of the first flare 
and extending until the burst position exited the LAT FoV or became occulted by the Earth (up to 1\,ks duration).

The searches were performed by means of an unbinned-likelihood analysis~\citep{080825c}. 
The searches over short time intervals ($<$400\,s) were performed using the Pass~7  Transient Class (``P7TRANSIENT'') data, 
appropriate for signal-limited analyses, and the relevant instrument response function P7TRANSIENT\_V6.
The searches over longer-duration intervals were performed  using the Pass~7 Source Class (``P7SOURCE'') data, 
appropriate for background-limited analyses, and the relevant instrument response function P7SOURCE\_V6 \citep{pass7}.

The analysis used events reconstructed within 15$^\circ$ around the XRT localization 
and with energies in the 100\,MeV--10\,GeV range. 
The X-ray flare spectrum was modeled using a power law with a free normalization and spectral index. 
The residual cosmic-ray background and the extragalactic gamma-ray background (CREGB) for the P7TRANSIENT analysis 
was estimated following the method described in \citet{080825c,vlasios13}. 
The Galactic diffuse background for both the P7TRANSIENT and P7SOURCE analyses, and the CREGB for the P7SOURCE analyses 
were modeled using the standard, publicly distributed, templates 
\footnote{gal\_2yearp7v6\_v0.fits, and iso\_p7v6source.txt available at
http://fermi.gsfc.nasa.gov/ssc/data/access/lat/BackgroundModels.html}. 
The background contribution from the Earth's atmospheric gamma-rays was negligible 
as the GRB positions were far from the Earth's limb during all the time intervals analyzed.
No point source in the vicinity of any of the analyzed GRBs (within 15$^\circ$) was bright enough to merit inclusion in the background model. 
In the case of a stacked analysis, we simultaneously fit the likelihood functions 
of the flares under search. A single flux and spectral index were fit to the stacked data.

High-energy emission in coincidence with the observed X-ray flares 
was detected in the case of GRB 100728A, 
as reported in \citet{100728a}. 
For the other bursts in our sample we report in Table~\ref{latul}
the upper limits in the 100\,MeV--10\,GeV energy range.
The quoted values are at a 95\% confidence level, and were calculated
for a photon index $\Gamma$=2.0.
The upper limits calculation used a Bayesian approach with a flat prior, in which the profile likelihood 
was treated as the posterior probability of the source flux, as described in Section~\ref{sec:gbm}.
The typical upper limit derived during each flare is an order of magnitude higher than the
flux measured in the case of GRB~100728A. In the case of multiple flares, the upper limits derived 
from the stacked analysis are a few times higher.
Two are the main differences between the flares in
our sample and those in GRB~100728A: 1) GRB~100728A triggered an autonomuous repointing of the instrument, so that its emission during the
flares was observed nearly on-axis where the sensitivity is maximum.
All the flares in our sample were instead serendipitously 
observed by the LAT, mostly at large off-axis angles; 2) GRB~100728A displayed an unusual series of
multiple, bright flares, and a detection was achieved only by considering the entire period of emission.
Most bursts in our sample display instead one or two flares.
Our results show that in general X-ray flares are not accompanied by a bright counterpart in the 
MeV-GeV energy range. The detection of a high-energy counterpart in GRB~100728A was the result of a
fortunate combination of sensitive observations and long-lived flaring activity.

\begin{table}[!t]
 \caption{LAT Upper Limits}
\begin{tabular}{lccccc}
\hline 
GRB\ \ \ \ \ \ \ \ \ \ \ \  & \ \  Description \ \  &
\ \ \ \ \ \ $t_{\rm i}$\ \ \ \ \ \  & \ \ \ \ $t_{\rm f}$\ \ \ \  
&  \ \ \ $<$$F_{\gamma,UL}>$  \ \ \  \\
 (1) & (2) & (3) & (4) & (5)  \\
\hline 
080906\dotfill  & Flare \#1 & 160 & 240 & 13 \\
 		& Flare \#2 & 520 & 800 & 4.2 \\
 		& All Flares & -- & -- & 3.3 \\
 		& Extended Period & 160 & 1160 & 1.7 \\
080928\dotfill  & Flare & 340 & 400 & 13 \\
 		& Extended Period & 340 & 1000 & 1.4 \\
081203A \dotfill  & Flare & 98 & 152 & 12 \\
 		  & Extended Period & 98 & 1098 & 1.8  \\
090407\dotfill 	& Flare 	  & 120 & 170 	& 15  \\
 		& Extended Period & 120 & 650 	& 1.9 \\
090516\dotfill  & Flare & 260 & 315 & 30 \\
 		& Extended Period & 260 & 560 & 6.5 \\

090831C\dotfill & Flare 	  & 160 & 260 	& 15 \\
                & Extended Period & 160 & 1060 	& 1.7 \\		
091221\dotfill  & Flare 	  & 95 & 160 	& 15 \\
 		& Extended Period & 95 & 260 	& 6.9 \\
100212A\dotfill & Flare \#1 	  & 76 & 85  	& 50 \\
 		& Flare \#2 	  & 95 & 160 	& 23 \\
 		& Flare \#3 	  & 223 & 243 	& 34 \\
 		& Flare \#4 	  & 249 & 275 	& 40 \\
 		& Flare \#5 	  & 344 & 383 	& 17 \\
 		& Flare \#6 	  & 413 & 445 	& 21 \\
 		& Flare \#7 	  & 628 & 750 	& 8.2 \\
 		& All Flares	  & --  & --  	& 4.2 \\
 		& Extended Period & 76  & 976 	& 1.4 \\
100725B\dotfill & Flare \#1 & 115 & 150 & 17 \\
 		& Flare \#2 & 150 & 200 & 15 \\
 		& Flare \#3 & 200 & 255 & 9.9 \\
 		& Flare \#4 & 255 & 315 & 1.2 \\
 		& All Flares & -- & -- & 4.2 \\
 		& Extended Period & 115 & 1115 & 0.9  \\
110102A\dotfill & Flare \#1 & 196 & 248 & 17 \\
        	& Flare \#2 & 249 & 393 & 4.5 \\
        	& All Flares & -- & -- & 3.5 \\
        	& Extended Period & 196 & 806 & 2.0 \\
110414A\footnote{The GRB position became occulted by the Earth at t=690 s. No search for extended high-energy emission was possible.}
\dotfill	& Flare  & 271 & 690 & 17 \\ 
110731A\footnote{High-energy emission is detected up to 1000~s \citep{110731a}.
Here we report the results only during the X-ray flare time interval.}
\dotfill& Flare  & 55 & 95 & 21 \\
\hline
\label{latul}
\end{tabular}
{\sc Notes :} Col. (1): GRB name;
Col. (2): We searched for high-energy emission in coincidence with each flare, 
by stacking the emission of multiple flares (All flares), and over
longer timescales (Extended Period);
Col.~(3): start time of the search, in units of s;
Col.~(4): stop time of the search, in units of s;
Col. (5): 95\% upper limit in the 100 MeV-10 GeV band. Units are \e{-9}\flux.
\end{table}

\newpage
\section{Theory and model description}\label{sec:model}
We developed analytical prescriptions for the synchrotron, and 
first-order Inverse Compton components that 
include self-absorption, opacity for pair
production and Thomson scattering on pairs, Klein-Nishina
effects, and the maximum energy for acceleration. The model covers
the various ordering of synchrotron self-absorption, maximum and
cooling frequencies. The model was implemented for use within XSPEC for
broadband spectral fitting.  XSPEC allows for
a simultaneous fit to all data sets by performing a minimization on
the PG-STAT statistic on LAT data, and on $\chi^2$ on the data sets at
lower energies. The use of PG-STAT is required by the low counts of LAT spectra, 
which are therefore Poissonian, and by the Gaussian uncertainties of 
the estimated LAT background \citep{vlasios13}. 
In this case, neither the $\chi^2$ nor the Cash statistics 
yield accurate results \citep{arnaud11}. 
Spectra and response matrices for the {\it Fermi} data were 
created following \citet{latcat}.

 The spectrum produced by an ensemble of electrons
accelerated by internal shocks is uniquely determined by six
parameters \citep{sapina98,piran04}, namely the internal energy  of the shock $e'$ (where
primed quantities are in the rest frame of the shocked fluid), the fraction
of energy that goes to electrons $\epsilon_e$ and to magnetic
field $\epsilon_B$, the electron spectral index $p$,
the bulk Lorentz factor\footnote{We assumed a contrast in the Lorentz
factors of the colliding shells
$\Delta \Gamma$\,$\approx$\,$\Gamma$. Smaller values are disfavored 
by the observations \citep{krimm07}.}
$\Gamma$, and the radius of the source $R$.
In the regime of fast cooling, which is the case
relevant to our observations, it is more convenient to 
replace $e'$ and $R$ with two related parameters more directly
linked to observable quantities \citep{gg03}. The first is the isotropic
electron luminosity $L_e$=$\frac{4}{3} \pi R^2 c \Gamma^2 \epsilon_e
e'$ which is equal, for fast cooling, to the total (synchrotron plus higher
IC orders) radiated luminosity $L_{\rm ISO}$=$L_{52}$10$^{52}$~erg\,s$^{-1}$. The other is the variability time
scale of the relativistic flow $t_v$,
which in the internal shock model is
related to the radius of the collision between shells by $R\approx2\Gamma^2 c t_v$.

Here we first briefly derive the characteristic frequencies of a synchrotron spectrum
according to our formalism.  Each electron
radiates a power $P(\gamma)$ at a typical synchrotron frequency $\nu_s$:
\begin{equation}
P(\gamma)=\frac{4}{3} c \sigma_T \gamma^2 \frac{B'^2}{8\pi}\Gamma^2  \label{psyn}
\end{equation}
\begin{equation}
\nu_s = \frac{3}{16}\frac{q_e B'}{m_e c} \gamma^2 \Gamma \label{nusyn}
\end{equation}
where $B'$ is the magnetic field, $\gamma$ the electron Lorentz factor, 
$m_e$ and $q_e$ its mass and electric charge respectively. 
The spectrum of the shock accelerated electrons is usually described as a power law
with index $p$, $dn'/d\gamma \propto \gamma^{-p}$ for $\gamma$$\geq$$\gamma_m$. 
We hardcoded a lower limit $p$$>2$ in the fitting routine in order to avoid 
divergent electron energies.
The minimum Lorentz factor $\gamma_m$, and the corresponding frequency $\nu_m$ (from Eq.~\ref{nusyn}) are given by:
\begin{subequations}
\begin{eqnarray} \label{gum}
\gamma_m&=&\epsilon_e \frac{m_p}{m_e} \frac{p-2}{p-1} \\
\nu_m&=&2.6 \times 10^{18}\ \left(\frac{p-2}{p-1}\right)^2 \epsilon_e^{3/2} \epsilon_B^{1/2} L_{52}^{1/2}
\Gamma_{2.5}^{-2} t_v^{-1}\ {\rm Hz\ \ \ \ \ \ \ } \label{num}
\end{eqnarray}
\end{subequations}
where $\Gamma_{2.5}=\Gamma/10^{2.5}$.
The time scale for radiative cooling by synchrotron and IC becomes
lower than the dynamical time scale above the cooling Lorentz
factor $\gamma_c$ at the frequency $\nu_c$:
\begin{subequations}
\begin{eqnarray}\label{nuc}
\gamma_c&=&2 (1+Y)^{-1} \epsilon_e \epsilon_B^{-1} L_{52}^{-1} \Gamma_{2.5}^5 t_v \\
\nu_c&=&3.7\times 10^{12} (1+Y)^{-2} \epsilon_e^{3/2} \epsilon_B^{-3/2} L_{52}^{-3/2}
\Gamma_{2.5}^{8} t_v {\rm Hz\ \ \ \ \ \ \ \ }
\end{eqnarray}
\end{subequations}
where $Y$ is the Compton parameter. In the limit of single scattering, it can be approximated 
by the ratio of luminosities radiated by IC and synchrotron:
\begin{equation}\label{Y}
Y=\frac{L_{IC}}{L_S}=\frac{-1+ \sqrt{1+4\eta \frac{\epsilon_e}{\epsilon_B} }}{2}
\end{equation}
where the radiation efficiency $\eta\approx1$ in the fast cooling regime.
Likewise, the maximum Lorentz factor $\gamma_M$ of the electrons is
derived by equating the acceleration time, which is essentially
the Larmor time, with the dynamical time scale \citep{piran04}:
\begin{subequations}
\begin{eqnarray}\label{nuM}
\gamma_M&=& 1.7 \times 10^6 (1+Y)^{-1/2} \epsilon_e^{1/4} \epsilon_B^{-1/4} L_{52}^{-1/4} \Gamma_{2.5}^{3/2} t_v^{1/2} \\
\nu_M&=&2.3\times 10^{24} (1+Y)^{-1} \Gamma_{2.5} {\rm Hz\ \ \ \ \ \ \ \ \ \ \ \ \ \ \ \ \ \ \ \ \ \ \ \ \ \ \
\ }
\end{eqnarray}
\end{subequations}
where $\nu_M$ is the maximum synchrotron frequency.

We derived the spectrum for various ordering of the
frequencies and extended the model prescriptions to the case of
slow cooling. This has been necessary because the fitting
procedure explores a wide range of parameters, often outside the
region of fast cooling. Below we summarize the case
$\nu_c<\nu_{sa}<\nu_{m}$ which applies to all the best-fit models.
Specifically, in the fast cooling regime, the spectrum of
electrons at equilibrium is given by:
\begin{equation}\label{eldistr}
N(\gamma)= N(\gamma_c)
\begin{cases}
\left(\frac{\gamma}{\gamma_c}\right)^{-2}  &\gamma_c<\gamma<\gamma_m \\
\left(\frac{\gamma}{\gamma_c}\right)^{-p-1} &\gamma_m<\gamma<\gamma_M\\
\end{cases}
\end{equation}
where $p$ is the slope of the injection spectrum.
Above $\nu>\nu_{\rm min}={\rm min}(\nu_m,\nu_c)$ the synchrotron spectrum
can be derived by adopting the delta function approximation, i.e. assuming
that all the power is radiated at the frequency given by Eq.~\ref{nusyn}. Then
 $L(\nu)d\nu=P(\gamma)N(\gamma) d\gamma$, using Eq.~\ref{psyn} and Eq.~\ref{eldistr},
yields:
\begin{equation}\label{ls}
L(\nu)= f_{\rm max} \left(\frac{\nu}{\nu_c}\right) ^{-\frac{q-1}{2}}
\end{equation}
 where $q$=2 for $\gamma_c$$<$$\gamma$$<$$\gamma_m$ and $q$=$p+1$ for
$\gamma_m$$<$$\gamma$$<$$\gamma_M$. The monochromatic luminosity $L_{\nu}$ peaks
at $\nu_c$ and is \mbox{$f_{\rm max}= m_e c^2 \sigma_T \Gamma B' N_e / 6 e$} where the total number of electrons $N_e=\gamma_c
N(\gamma_c)$ for $\gamma_c<<\gamma_m$.

At low frequencies the effect of synchrotron self-absorption
becomes relevant. The absorption coefficient $\alpha_{\nu}$ can be
specified \citep{pk00} as:
\begin{equation}
\label{alphasa}
\alpha_{\nu} \propto \begin{cases} \nu^{-5/3}, & \mbox{for } \nu<\nu_c \\
\nu^{-3}, & \mbox{for } \nu_c<\nu<\nu_m
\end{cases}
\end{equation}
and the spectrum can be derived recalling that, in the homogeneous
case, the intensity in the optically thick regime is proportional
to the source function \citep{rybicki}:
\begin{equation}\label{lsa}
L_\nu\propto \frac{P(\nu)}{\alpha_{\nu}}\propto
\begin{cases}
\nu^{2}, & \nu<\nu_c \\
\nu^{5/2}, & \nu_c<\nu<\nu_{sa}
\end{cases}
\end{equation}
The synchrotron self absorption frequency $\nu_{sa}$ is derived by requiring
that the optically thick emission (Eq.~\ref{lsa}) equals the the optically
thin emission (Eq.~\ref{ls}):
\begin{equation}\label{nusa}
\nu_{sa}=9.2 \times 10^{14} (1+Y)^{-1/3} \epsilon_e^{-1/3}  L_{52}^{1/3}
\Gamma_{2.5}^{-1} t_v^{-2/3} {\rm \ Hz\ \ \ \ }
\end{equation}

Summarizing the above equations, the synchrotron spectrum is:
\begin{equation}
\label{syncspe}
\frac{\nu L_{\nu}}{\nu_m L_{\nu_m}}=
\begin{cases}
\left(\frac{\nu_{sa}}{\nu_m}\right)^{1/2}\left(\frac{\nu_c}{\nu_{sa}}\right)^{7/2} \left(\frac{\nu}{\nu_c}\right)^3, &\nu < \nu_c  \\
\left(\frac{\nu_{sa}}{\nu_m}\right)^{1/2}\left(\frac{\nu}{\nu_{sa}}\right)^{7/2} , &\nu_c < \nu < \nu_{sa} \\
\left(\frac{\nu}{\nu_m}\right)^{1/2} , &\nu_{sa} < \nu < \nu_m \\
\left(\frac{\nu}{\nu_m}\right)^{1-p/2} , &\nu_m < \nu < \nu_M
\end{cases}
\end{equation}
with the normalization
\begin{equation}
\label{snorm}
\nu_m L_{\nu_m}= 5\times 10^{51}
\left(\frac{p-2}{p-1}\right) (1+Y)^{-1} L_{52} \mbox{~erg\,s$^{-1}$}
\end{equation}
We also considered the case where the electron distribution
is inhomogeneous \citep{gps00}. In this case a new
absoption frequency $\nu_{ac}$$<$$\nu_{sa}$ appears. The spectrum is
modified as $L_\nu$\,$\propto$\,$\nu^{11/8}$ for $\nu_{ac}$$<$$\nu
<\nu_{sa}$, and $\L_\nu$\,$\propto$\,$\nu^2$ for $\nu$$<$$\nu_{ac}$ \citep{gg03}.

The IC spectrum has been derived following the prescriptions of 
\citet{sariesin01}.  In the Thomson limit, the energy of the upscattered photon is
$\nu=4 x \gamma^2 \nu_s$, with $0<x\leq 1$ and the cross section is approximated by the Thompson value. The spectrum is then given by
\begin{equation}\label{LIC}
L_{\nu}^{IC}=R \sigma_T \int_{\gamma_{min}}^\infty d\gamma N(\gamma) \int_0^1 dx\  g(x) L_{\nu_s}(x)
\end{equation}

where $\gamma_{\rm min}={\rm min}(\gamma_c, \gamma_m)$ and the Green function $g(x)$ \citep{bg70}  gives
the probability of producing an IC photon at  frequency $\nu$  from a synchrotron photon at $\nu_s$. Following \citet{sariesin01}, we approximated $g(x)$$\sim$1 for $0<x<0.5$
The inner integral of Eq.~\ref{LIC} represents the IC spectrum produced by monoenergetic electrons. For a
given power-law segment with $f(\nu_s)\propto\nu_s^{\alpha}$, the integral becomes
$I\propto\nu^{\alpha}$ for  $\alpha < 1$ and
$I\propto\nu^{1}$ for  $\alpha > 1$.
Thus the IC spectrum of monoenergetic electrons has the same shape above $\gamma^2\nu_{sa}$ as the
input synchrotron spectrum above $\nu_{sa}$, with all frequencies  boosted by a factor $\gamma^2$. Below $\gamma^2\nu_{sa}$ the effect of the distribution dominates and the integral is linear.
For the case under examination   (Eq.~\ref{syncspe}) the spectrum is:
\begin{equation}
\label{Ispe}
I\propto f_{max}
\begin{cases}
\left(\frac{\nu_{sa}}{\nu_c} \right)^{-1/2}\left(\frac{\nu}{\gamma^2\nu_{sa}} \right), & \nu < \gamma^2 \nu_{sa} \\
\left(\frac{\nu_m}{\nu_c} \right)^{-1/2}\left(\frac{\nu}{\gamma^2\nu_m} \right)^{-1/2}, & \gamma^2 \nu_{sa} <\nu < \gamma^2 \nu_m\\
\left(\frac{\nu_m}{\nu_c} \right)^{-1/2}\left(\frac{\nu}{\gamma^2\nu_m} \right)^{-p/2}, & \gamma^2 \nu_m <\nu \\
\end{cases}
\end{equation}
Note that  the  shape of the IC  spectrum below $\gamma^2 \nu_{sa}$  is the same for either the homogenous
and inhomogeneous cases because below $\nu_{sa}$  the corresponding power-law segments have indices  greater than 1.
Substituting Eq.~\ref{Ispe} in  Eq.~{\ref{LIC} gives the approximate solution:
\begin{equation}
\label{icspe}
\nu L^{IC}_{\nu} \approx \nu_m^{IC} L_{\nu_m}^{IC}
\begin{cases}
\left(\frac{\nu_{sa}^{IC}}{\nu_m^{IC}}\right)^{1/2} \left(\frac{\nu}{\nu_{sa}^{IC}}\right)^2, &\nu < \nu_{sa}^{IC}  \\
\left(\frac{\nu}{\nu_m^{IC}}\right)^{1/2} , &\nu_{sa}^{IC}< \nu < \nu_{m}^{IC} \\
\left(\frac{\nu}{\nu_m^{IC}}\right)^{1-p/2} , &\nu_{m}^{IC} < \nu < \nu_M^{IC} \\
\end{cases}
\end{equation}
where
\begin{eqnarray}
\nonumber
\nu_m^{IC} L_{\nu_m}^{IC} &=&  Y \nu_m L_{\nu_m}\\
\nonumber
\nu_{sa}^{IC}  &=&   \gamma^2_c \nu_{sa} \\
\nonumber
\nu_m^{IC}  &=&  \gamma_m^2 \nu_m  \\
\nu_M^{IC}  &=&  \gamma_M^2 \nu_M
\end{eqnarray}
and the corresponding numerical values are obtained by substituting Eq.~\ref{num},~\ref{nuc},
\ref{nuM},~\ref{nusa},~\ref{snorm}.
The spectral shape given in Eq.~\ref{icspe} was derived by assuming the IC scattering 
to take place in the Thomson regime, i.e., when the $\gamma h \nu_s<\Gamma mc^2$. Electrons with $\gamma> \tilde{\gamma}_m=\frac{mc^2\Gamma}{h\nu_m}$ interact with  photons with $\nu>\nu_m$ in the Klein-Nishina regime. 
In this regime, the upscattered energy is $h \nu^{IC}= \Gamma \gamma mc^2$.
Thus the transition appears in the IC spectrum at:
\begin{equation}
h \nu_{KN}^{IC}  =  \tilde{\gamma}_m mc^2 \Gamma= \frac{(mc^2)^2 \Gamma^2}{h\nu_m} \\
\end{equation}
when $\nu_{KN}^{IC}< \nu_M^{IC}$ with
\begin{equation}
\label{nuicmax}
\nu_M^{IC}=\frac{\gamma_M mc^2}{h}\Gamma
 \end{equation}
In this energy range,  the spectrum can be approximated by $L_{\nu}^{IC}\propto \nu^{-(p+1)/2}$ \citep{gg03}.
\citet{nakar09} presented a more detailed calculation of the Klein-Nishina
regime, showing that the spectral shape could be significantly modified with respect to the simple treatment 
of \citet{gg03}. However, as in our case the peak of the synchrotron emission is in the soft X-ray range, 
Klein-Nishina effects can be considered negligible, and the adopted approximation does not affect our results. 
In fact, in fast cooling, the Klein-Nishina effects can be neglected when $\gamma_m<<{~\tilde{\gamma}}_m$. 
By substituting Eq.~\ref{num}, we derive the condition $\Gamma>> \epsilon_e \frac{h\nu_m}{1 \rm keV}$, 
which is always satisfied for the kind of relativistic flow that we consider here.

In the low energy part of the spectrum, the IC photons are also
subject to synchrotron absorption at $\nu$$<$$\nu_{sa}$, when the
source becomes optically thick. In this regime, the spectrum is
proportional to the source function (Eq.~\ref{lsa}) which, for the
absorption coefficient given in Eq.~\ref{alphasa}, yields
$L_{\nu}^{IC}\propto \nu^{4}$.  The overall IC spectrum is then:
\begin{equation}
\label{icspe}
\frac{\nu L^{IC}_{\nu}}{\nu_m^{IC} L_{\nu_m}^{IC}}=
\begin{cases}
\left(\frac{\nu_{sa}^{IC}}{\nu_m^{IC}}\right)^{1/2} \left(\frac{\nu_{sa}}{\nu_{sa}^{IC}}\right)^2 \left(\frac{\nu}{\nu_{sa}}\right)^5, &\nu < \nu_{sa} \\
\left(\frac{\nu_{sa}^{IC}}{\nu_m^{IC}}\right)^{1/2} \left(\frac{\nu}{\nu_{sa}^{IC}}\right)^2, &\nu_{sa}<\nu < \nu_{sa}^{IC}  \\
\left(\frac{\nu}{\nu_m^{IC}}\right)^{1/2} , &\nu_{sa}^{IC}< \nu < \nu_{m}^{IC} \\
\left(\frac{\nu}{\nu_m^{IC}}\right)^{1-p/2} , &\nu_{m}^{IC} < \nu < \nu_{KN}^{IC} \\
\left(\frac{\nu_{IC}^{KN}}{\nu_m^{IC}}\right)^{1-p/2}\left(\frac{\nu}{\nu_{KN}^{IC}}\right)^{1/2-p}, &\nu_{KN}^{IC} < \nu < \nu_M^{IC}
\end{cases}
\end{equation}

In the high energy part of the spectrum, photons with energy $h
\nu_{\gamma \gamma}$ can interact with photons at energies $h
\nu_{an}$\,$\geq$\,${(\Gamma mc^2)^2}/{h \nu_{\gamma \gamma}}$ and
annihilate in pairs. In the case under study, by requiring that the optical depth for pair
production $\tau_{\gamma \gamma}$ is less than unity \citep{ls01} 
one obtains a spectral cut-off at:
\begin{equation}\label{egg}
\nu_{\gamma \gamma}= 5.6\times 10^{25} (1+Y)^{\frac{2}{p}}
(\epsilon_e^3\epsilon_B)^{\frac{2-p}{2p}}
L_{52}^{(\frac{p+2}{2p})}
 t_v \Gamma_{2.5}^{\frac{4(p+1)}{p}}
{\rm Hz}
\end{equation}

Finally, the optical depth for Thomson scattering on the pairs has
to be smaller than one. Taking into account that the number of
pairs is equal to the number of photons that annihilate, this
condition is satisfied when the cut-off energy for pair production
in the rest frame of the shell is greater than $mc^2$ \citep[e.g.][]{080825c}. By implementing the prescriptions of
\citet{ls01} we derive $\frac{h \nu_{\gamma
\gamma}}{\Gamma} \gtrsim 3 m c^2$, which gives a lower limit on the bulk Lorentz
factor $\Gamma$:
\begin{equation}\label{gammapp}
 \Gamma\gtrsim 220
 (1+Y)^{\frac{-2}{3p+4}}
(\epsilon_e^3\epsilon_B)^{\frac{p-2}{2(3p+4)}}
L_{52}^{\frac{p+2}{2(3p+4)}}
 t_{v,-2}^{\frac{-p}{3p+4}}
\end{equation}
Equations (\ref{egg},~\ref{gammapp}) apply to the case in which high energy photons annihilate on target photons dominated by the synchrotron component, that is the most common situation. 
In the IC case the cut-off energy can be significantly lower, modifying Eq.~\ref{egg} by a factor $Y^{-\frac{2}{p}} \gamma_m^{-\frac{2(p-2)}{p}}$. 
This is taken into account in the fitting program by implementing, 
in addition to the analytical approach, an iterative routine to avoid the 
regions of parameter space
that do not satisfy the opacity constraints.

It is useful for the discussion below  to rewrite the condition for Thomson opacity 
on pairs as an upper limit to $t_v$ in terms of two observable quantities, 
$\nu_m$ and  the flux $F_{\nu_m}$:
\begin{subnumcases}{t_v \lesssim}
 \nonumber 0.2 {\rm \ } 
 \left(1+Y\right)^{\frac{p+8}{2(p+4)}}  
 \left(\epsilon_{e,-1}^3\epsilon_{B,-1}\right)^{\frac{p+8}{2(p+4)}}  
 \left(\frac{h \nu_m}{{\rm 1\,keV}}\right)^{-\frac{3p+4}{p+4}}  \\  
 \left(\frac{\nu_m F_{\nu_m}}{10^{-9}~{\rm erg\,cm^{-2}\,s^{-1}}}\right)^{\frac{p}{2(p+4)}} 
 D_{28}^{\frac{p}{p+4}} \label{casoa}\\
 \nonumber \\
 \nonumber \\
 \nonumber 0.08 {\rm \ }  
 \left(\frac{Y}{10}\right)^{\frac{p}{2(p+4)}} 
 \epsilon_{e,-1}^{\frac{5(8-p)}{2(p+4)}} 
 \epsilon_{B,-1}^{\frac{(8+p)}{2(p+4)}}
 \left(\frac{h \nu_m}{{\rm keV}}\right)^{-\frac{3p+4}{p+4}} \\
 \left(\frac{\nu_m F_{\nu_m}}{10^{-9}~{\rm erg\,cm^{-2}\,s^{-1}}}\right)^{\frac{p}{2(p+4)}}  
 D_{28}^{\frac{p}{p+4}}  \label{casob}
\end{subnumcases}

Equation \ref{casoa} applies when the opacity is dominated by synchrotron. 
When the $Y$-parameter increases, the opacity is more likely
dominated by target photons on the IC component (Eq.~\ref{casob}). 
The numerical factor is computed for $p$=2.5 and
varies by less than 30\% for 2.5$<$$p$$<$3.

\section{Results}\label{sec:results}
\subsection{Constraints on IC processes}


Before presenting the results of the broadband fits, we discuss some general properties of the flares 
and their implications.
As reported in Table~1, X-ray flares are characterized by a peak energy 
within or below the X-ray band, a typical flux 
$f_X \approx 10^{-9}-10^{-8}$\,erg\,cm$^{-2}$\,s$^{-1}$,
and an observed duration $\Delta t$$\approx$10-100~s.
In agreement with previous studies on X-ray flares \citep{falcone07}, we found that $\sim$40\% of the spectra 
are curved and can be described by a Band function with $\alpha$$\approx$1.0
and $\beta$$\approx$2.4, and a peak energy of a few keV. 
In the remaining cases a power law with photon index $\Gamma$$\approx$2.2 is already a good fit,
suggesting that the spectral peak is below the XRT energy band.

A common feature of the flares in our sample is that usually no bright optical counterpart 
is observed during the time interval of the X-ray flares,
despite the good and simultaneous sampling with the UVOT (see Figure~1). 
The optical afterglow, when detected, 
does not show in general significant variability,
suggesting that emission from the forward shock is likely the dominant component. 
The optical counterpart of X-ray flares remains hidden within the measurement uncertainty,
which we adopt as an upper limit to the optical flare emission.
 
For a typical X-ray flux $\gtrsim$\e{-9}\,erg\,cm$^{-2}$\,s$^{-1}$, the optical 
non-detection implies an optical-to-X-ray spectral index $\beta_{ox}<0.5$.  
Such a flat spectrum could be due to a significant absorption affecting the UV-optical range. 
However it is unlikely that external factors (dusty environment, or high redshift) contribute
to the suppression of optical emission in the whole sample, and they can be ruled out for those 
GRBs with a measured redshift and intrinsic extinction.
The absence of optical flares is more likely an intrinsic property of the spectra of the X-ray flares, 
which could be explained either if the synchrotron self-absorption frequency $\nu_{sa}$ 
is above the optical band, or if the X-ray flare is produced by the IC scattering of optical/infrared 
synchrotron photons \citep{koba07}.
In the latter case, the lack of an optical counterpart implies that the low-energy synchrotron component 
is suppressed by an efficient ($Y$$>>$1) IC cooling, and therefore a bright second-order IC component,
peaking at $\nu_{\gamma}$$\approx$10-100 MeV, should be visible.
The simultaneous {\it Fermi}-GBM and {\it Fermi}-LAT observations disfavor this scenario, as we discuss below. 
The X-ray-to-optical flux ratio can be used to set a lower limit on the $Y$ parameter, 
derived by combining Equations~13, and 18:
\begin{equation}
\label{ylower}
Y \gtrsim \left(\frac{\nu_{X} f_{X}}{\nu_{opt} f_{opt}}\right) 
\left( \frac{\nu_{opt} \nu_{\gamma}}{\nu_X^2}\right)^{1+\beta}
\end{equation}
where $\nu_{\gamma}$ is the peak frequency of the second-order IC component.
The second term in Equation~25 takes into account the fact that the synchrotron 
peak lies below the observed optical range, at a frequency $\nu_m$$\approx$$\nu_X^2/\nu_{\gamma}$.

As $\nu_{\gamma}$  lies either in the GBM or LAT 
energy band, the gamma-ray-to-X-ray flux ratio sets an upper limit on the $Y$ parameter: 
$Y \lesssim f_{\gamma} \nu_{\gamma}/ f_X \nu_{X}$.
Such a limit is valid in the case of optically thin emission, 
which for a mildly relativistic outflow, $\Gamma$$\gtrsim$5 (1+$z$),
is always satisfied at an energy of $\approx$10~MeV.

As shown in Fig.~\ref{fig:ypar}, the constraints derived from optical 
and gamma-ray observations are inconsistent for the bright flares 
($f_X$$>$\e{-8}\,erg\,cm$^{-2}$\,s$^{-1}$) in our sample:
 the lower limits derived from optical data require in most cases $Y$$>$100, 
while the upper limits derived from gamma-ray data [GBM (top panel), and LAT (bottom panel)] 
imply much lower values of the $Y$-parameter, not consistent with the
constraints derived from optical data.
Therefore, the X-ray emission is unlikely to be dominated by the IC component.


\begin{figure}[!t]
\centering
\includegraphics[angle=270,scale=0.32]{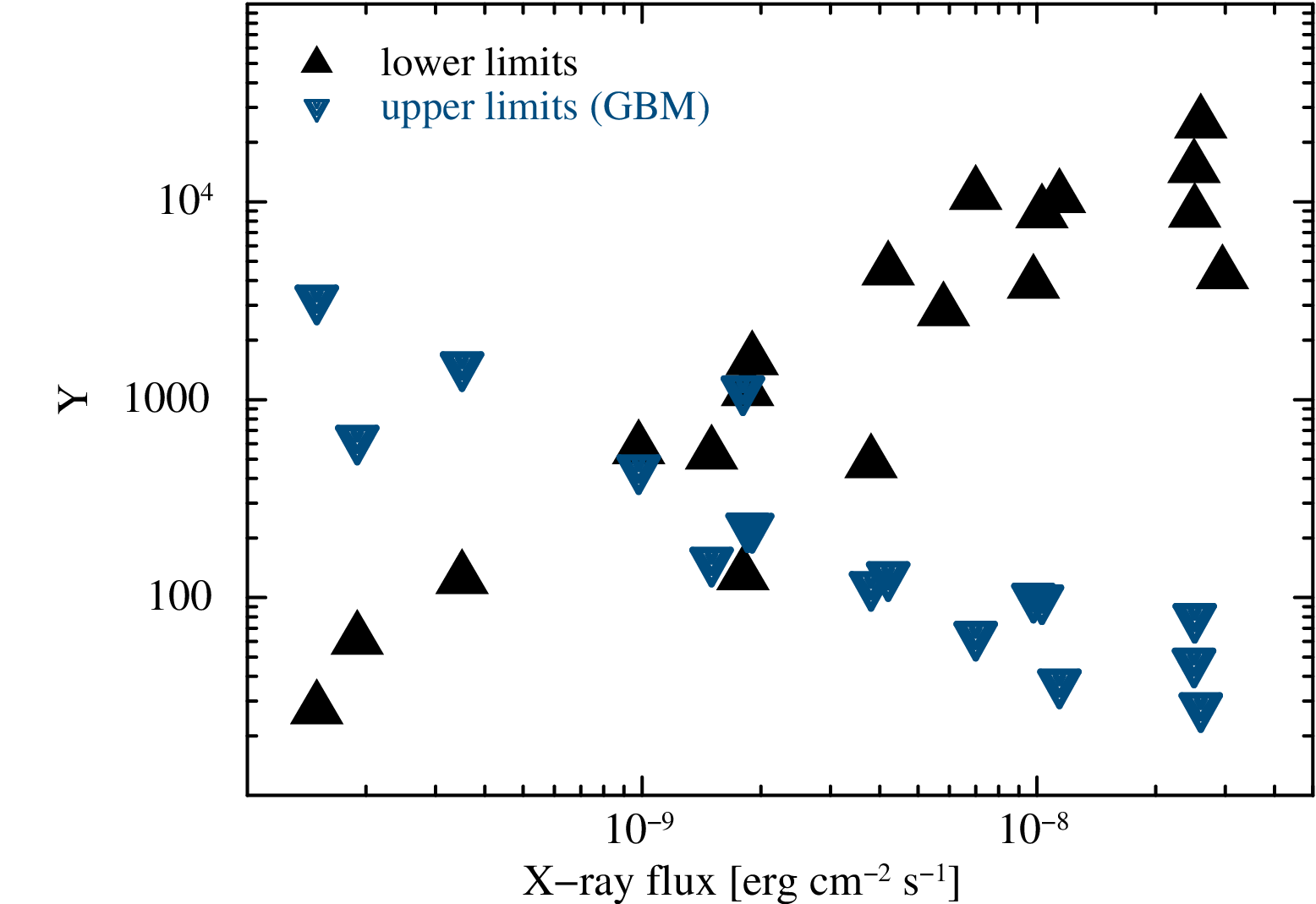} \\
\vspace{0.5cm}
\includegraphics[angle=270,scale=0.32]{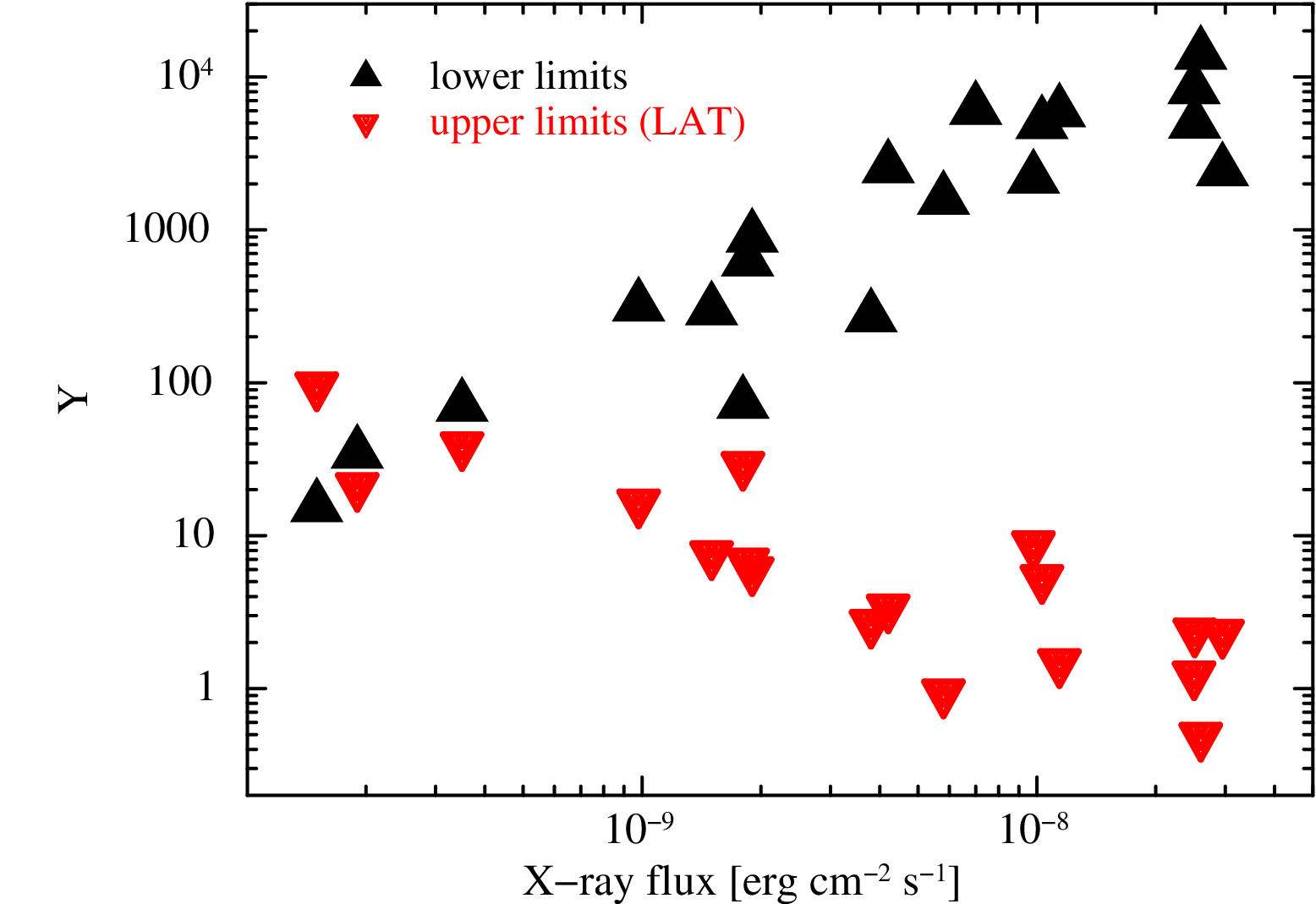}
\caption{{\it Top panel:} Constraints on the Y-parameter 
as a function of the observed X-ray flux.
The lower limits on the $Y$-parameter were calculated
according to Eq.~\ref{ylower} for an IC peak frequency $\nu_{\gamma}$=10~MeV. 
Upper limits were derived from {\it Fermi}/GBM observations. 
{\it Bottom panel:} Constraints to the Y-parameter 
as a function of the observed X-ray flux.
The lower limits on the $Y$-parameter were calculated
according to Eq.~\ref{ylower} for an IC peak frequency $\nu_{\gamma}$=100~MeV. 
Upper limits were derived from {\it Fermi}/LAT observations. 
}
\label{fig:ypar}
\end{figure}


\subsection{Broadband spectral modeling}

\begin{table*}
\begin{center}
\caption{Results of the broadband fit}\label{tab:fit}
\begin{tabular}{lccccccc}
\hline
GRB\ \ \ \ \ \ \ \ \ \ \ \ \ \ \ \ 
& \ \ \ \ \ \ \ $L_{52}$\ \ \ \ \ \ 
& \ \ \ \ \ \ \ $\epsilon_e$\ \ \ \ \ \ 
& \ \ \ \ \ \ \ $\epsilon_B$ \ \ \ \ \ \ 
& \ \ \ \ \ \ \ $p$\ \ \ \ \ \ 
& \ \ \ \ \ \ \ $t_v$\ \ \ \ \ \ 
& \ \ \ \ \ \ \ $\Gamma$\ \ \ \ \ \ 
& \ \ \ \ \ \ ${\rm STAT/dof}$\ \ \ \ \ \ \\
 (1) & (2) & (3) & (4) & (5) & (6) & (7) & (8) \\
\hline
080906\dotfill  & $0.05\pm0.02$  
		& $0.35^{+0.10}_{-0.3}$ 
		& $0.45^{+0.10}_{-0.15}$ 
		& 2.1$\pm$0.1  
		& 0.1 s 
		& 76$\pm$10
		& 113/94 \\
		& $0.11^{+0.12}_{-0.06}$  
		& 0.43$^{+0.10}_{-0.4}$
		& 0.07$^{+0.4}_{-0.05}$
		& 2.1$\pm$0.1 
		& 1 ms 
		& 270$^{+130}_{-50}$
		& 74/94 \\
080928\dotfill  & $0.012_{-0.002}^{+0.002}$  
		& $0.30_{-0.07}^{+0.10}$
		& $0.60_{-0.10}^{+0.10}$ 
		& 2.68$^{+0.10}_{-0.18}$ 
		& 0.1 s
		& $56_{-6}^{+10} $
		& 163/127\\
		
		& $0.020^{+0.05}_{-0.005}$  
		& $>0.08$
		& $>0.01$ 
		& $2.4^{+0.4}_{-0.2}$ 
		& 1 ms 
		& $400_{-200}^{+150} $
		& 164/127 \\
081203A\dotfill & 0.21$^{+0.17}_{-0.10}$  
		& 0.26$^{+0.12}_{-0.05}$ 
		& 1.4$^{+2.3}_{-0.9}$\ee{-2} 
		& 2.23$^{+0.3}_{-0.10}$ 
		& 1~ms
		& 300$^{+40}_{-50}$ 
		& 363/229 \\
090516\dotfill  & $0.26^{+0.12}_{-0.02}$  
		& $0.21_{-0.03}^{+0.12}$ 
		& $>0.25$ 
		& $3.2^{+0.3}_{-0.4}$ 
		& 0.1~s 
		& $100^{+25}_{-5} $
		& 157/157 \\
       		& $0.5_{-0.3}^{+0.5}$ 
		& $>0.02$ 
		& $>0.03$ 
		& $3.2_{-0.4}^{+0.3}$ 
		& 1~ms 
		& $>280$ 
		& 159/157\\
110731A\dotfill  & $0.43^{+0.16}_{-0.18}$  
		& $0.44_{-0.3}^{+0.11}$ 
		& $0.19_{-0.09}^{+0.11}$ 
		& $2.08^{+0.04}_{-0.07}$ 
		& 0.1~s 
		& 115$\pm$15
		& 120/104 \\
       		& $0.33_{-0.18}^{+0.2}$ 
		& $0.46_{-0.26}^{+0.04}$ 
		& $0.16_{-0.08}^{+0.16}$ 
		& 2.13$\pm$0.12 
		& 1~ms 
		& $450_{-200}^{+300}$ 
		& 104/104\\

\hline
\end{tabular}

\begin{minipage}[h]{0.83\linewidth}
\vspace{0.05in}
{\sc Notes :} Col. (1): GRB name;
Col.~(2): Bolometric luminosity (units are 10$^{52}$\,erg\,cm$^{-2}$\,s$^{-1}$);
Col.~(3): electron energy fraction;
Col.~(4): magnetic energy fraction;
Col.~(5): electrons spectral index;
Col.~(6): variability timescale;
Col.~(7): Lorentz factor;
Col.~(8): best fit statistic;
\end{minipage}
\end{center}
\end{table*}

\begin{figure*}[!t]
\centering
\includegraphics[angle=270,scale=0.31]{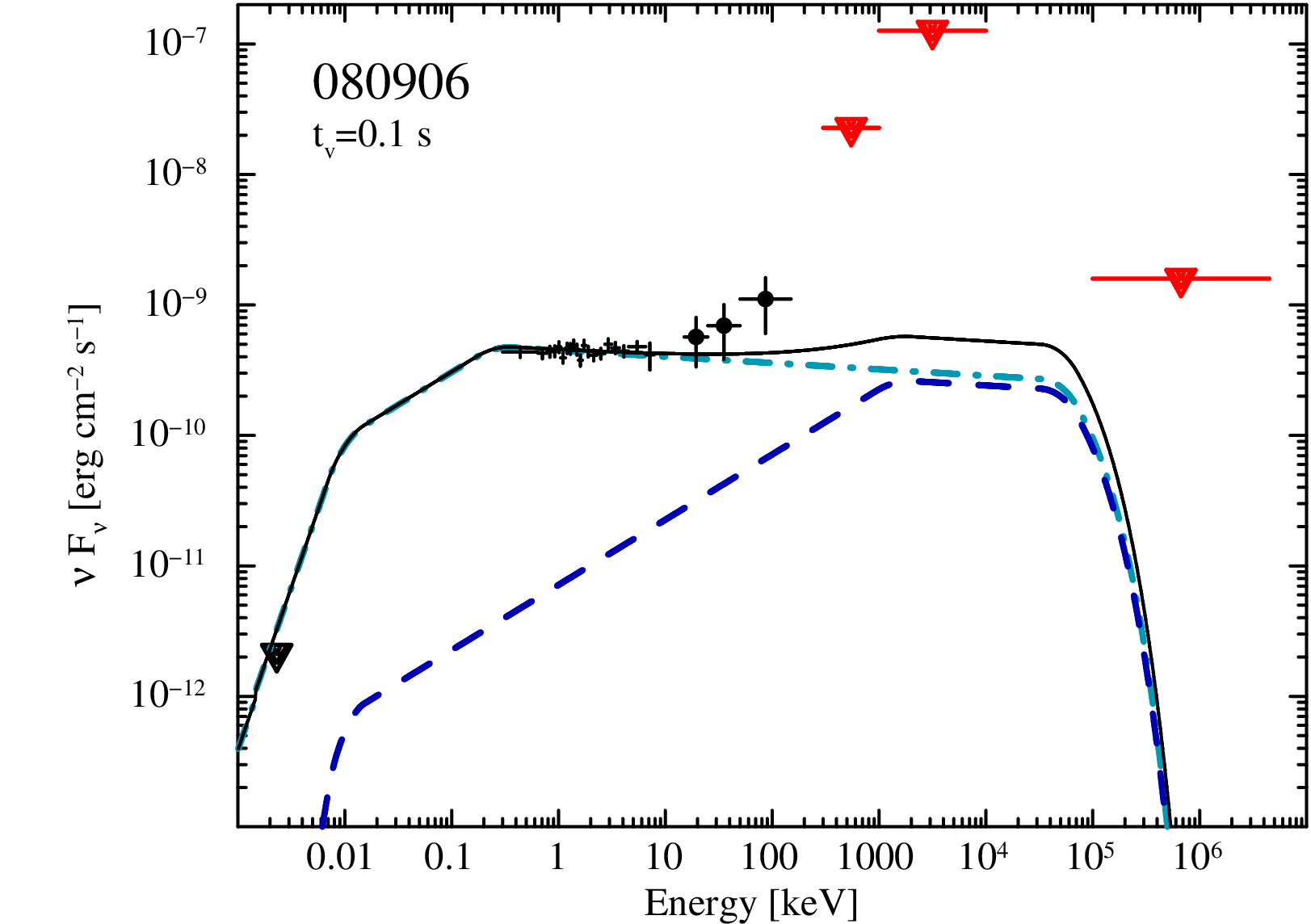}
\includegraphics[angle=270,scale=0.31]{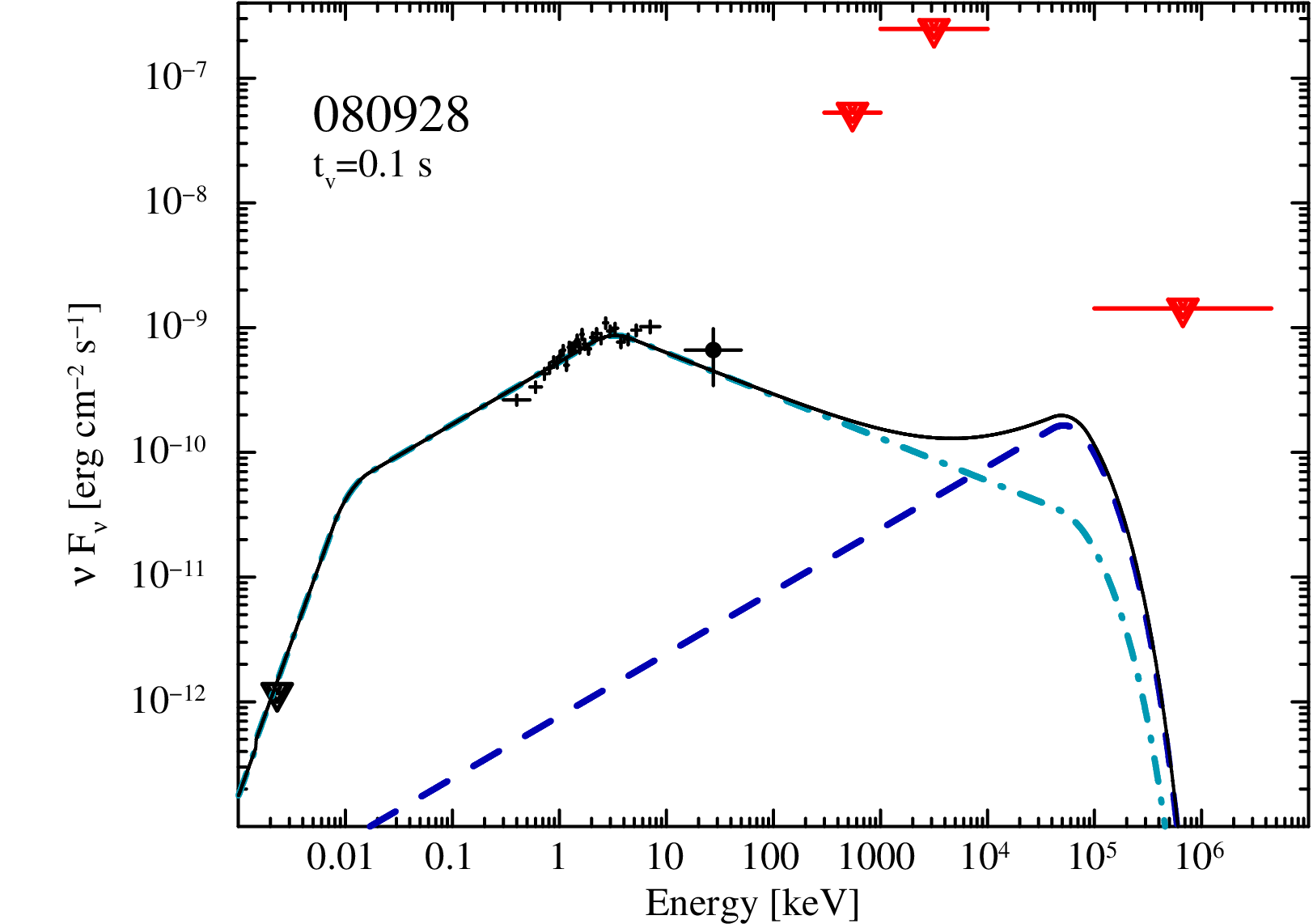}\\
\vspace{0.5cm}
\includegraphics[angle=270,scale=0.31]{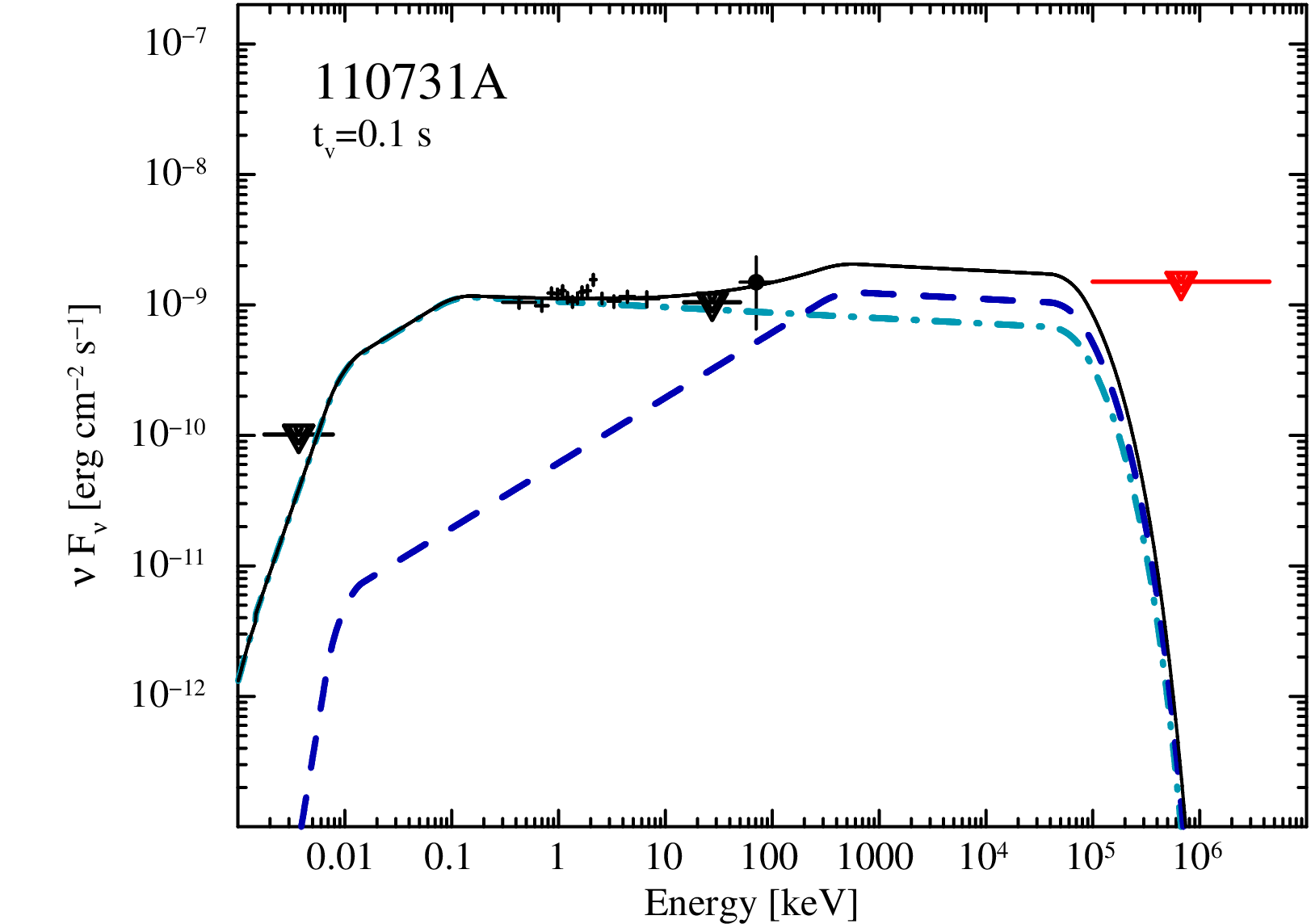}
\includegraphics[angle=270,scale=0.31]{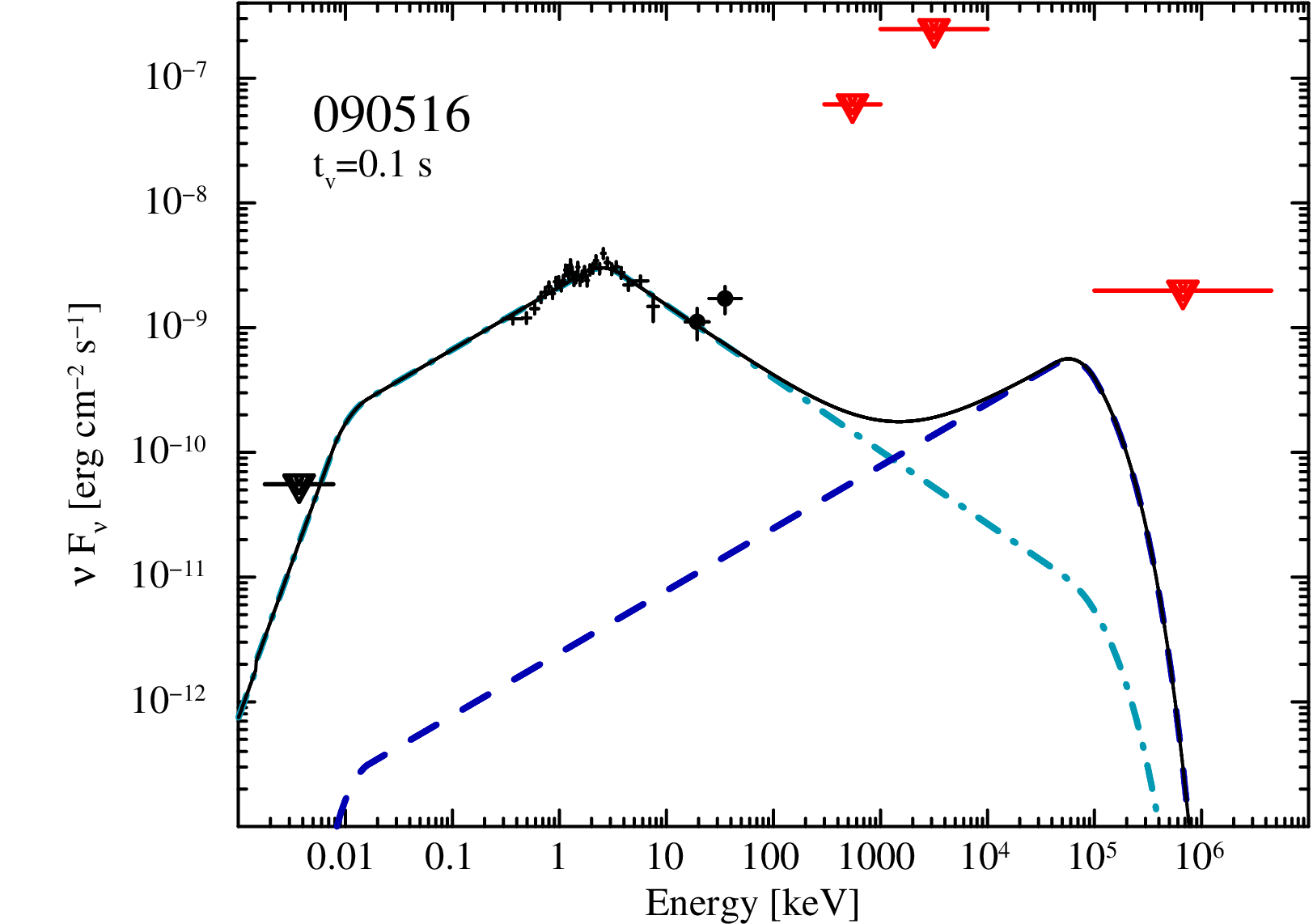}
 \caption{Broadband spectra of X-ray flares comparing \swift~(black) and 
 \fermi~(red) data with the standard internal shock model.
The best-fit model for $t_v$=0.1~s is shown by the solid line. 
The synchrotron component is shown by the dot-dashed line, 
the IC component by the dashed line.
The LAT upper limits were calculated in two energy bands, 
100 MeV - 3 GeV and 3 GeV - 30 GeV, 
using the procedure described in Section.~\ref{lat}, and converted into
energy flux units using the best-fit spectral model. Upper limits above 3 GeV
are out of scale, and are not shown.}
\vspace{0.5cm}
\label{bestfit}
\end{figure*}

\begin{figure*}[!t]
\centering
\includegraphics[angle=270,scale=0.31]{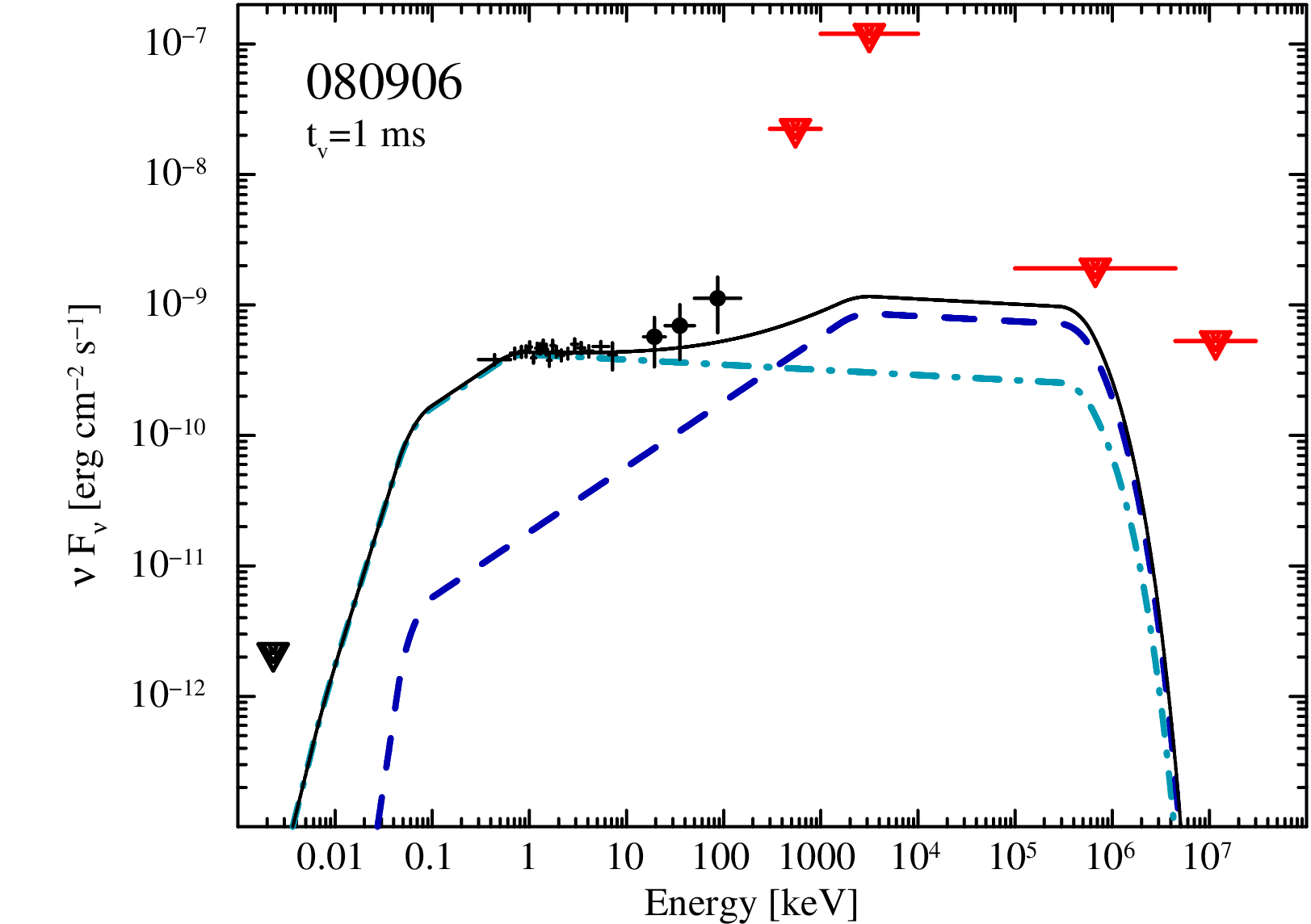}
\includegraphics[angle=270,scale=0.31]{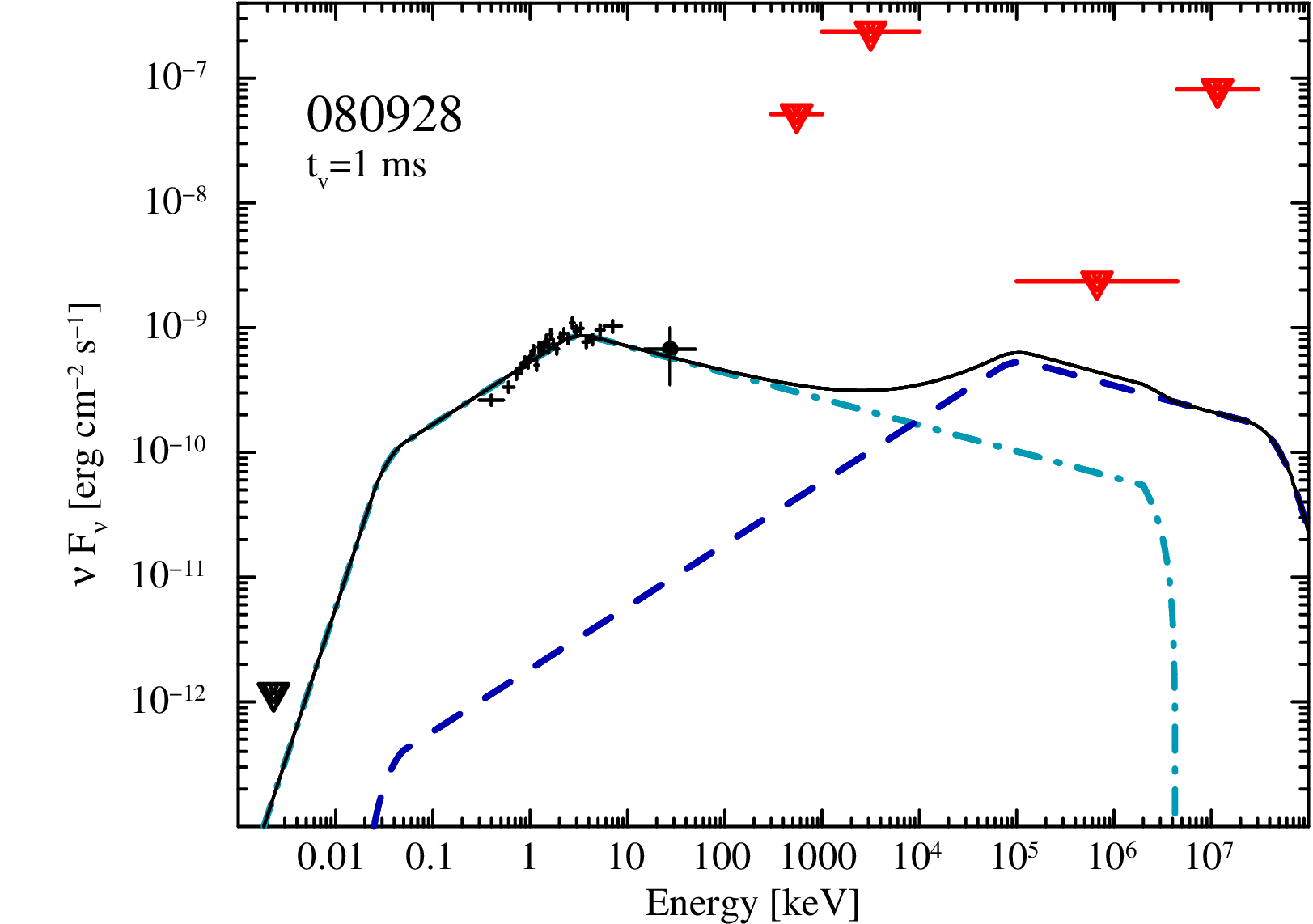}\\
\vspace{0.5cm}
\includegraphics[angle=270,scale=0.31]{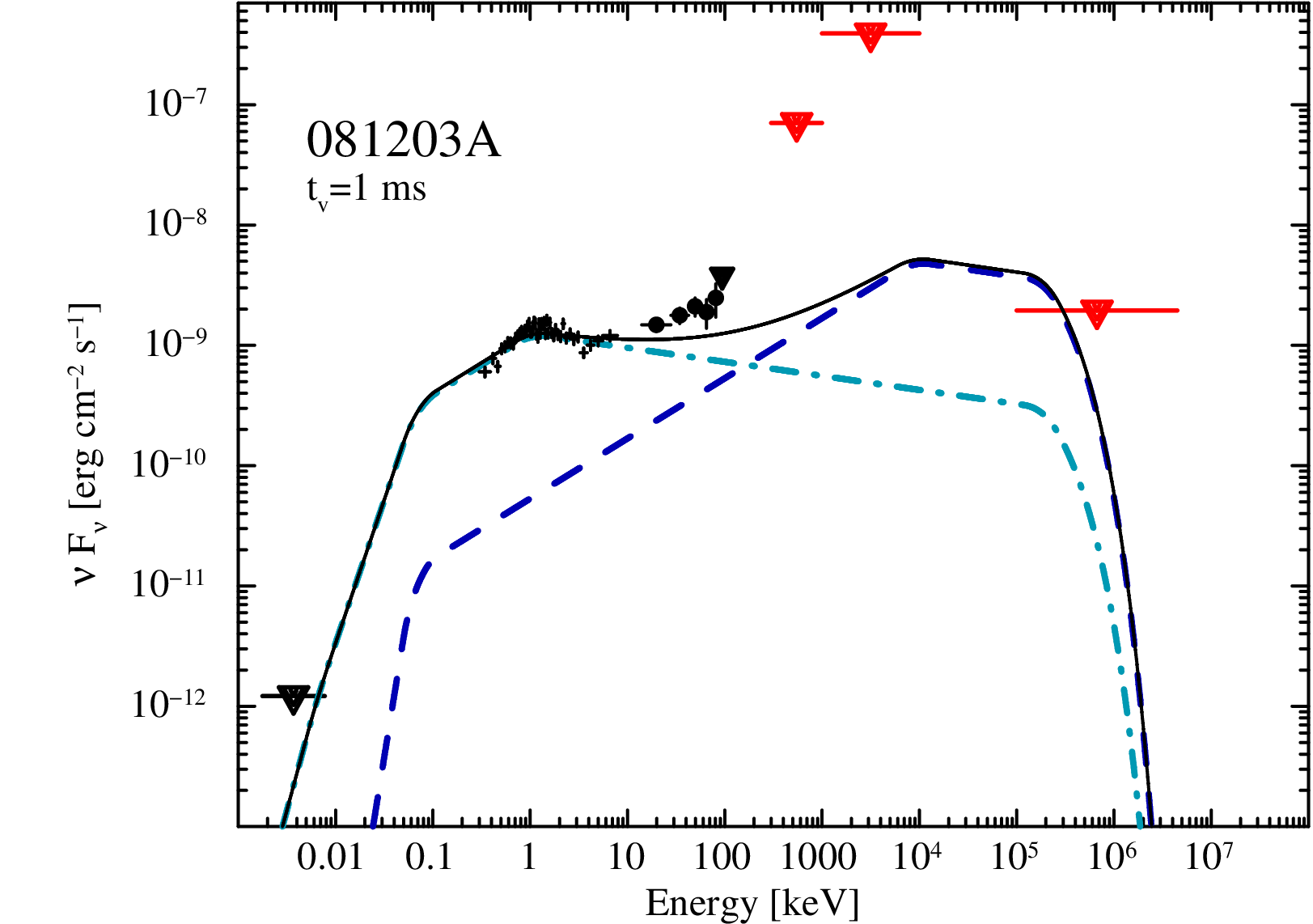}
\includegraphics[angle=270,scale=0.31]{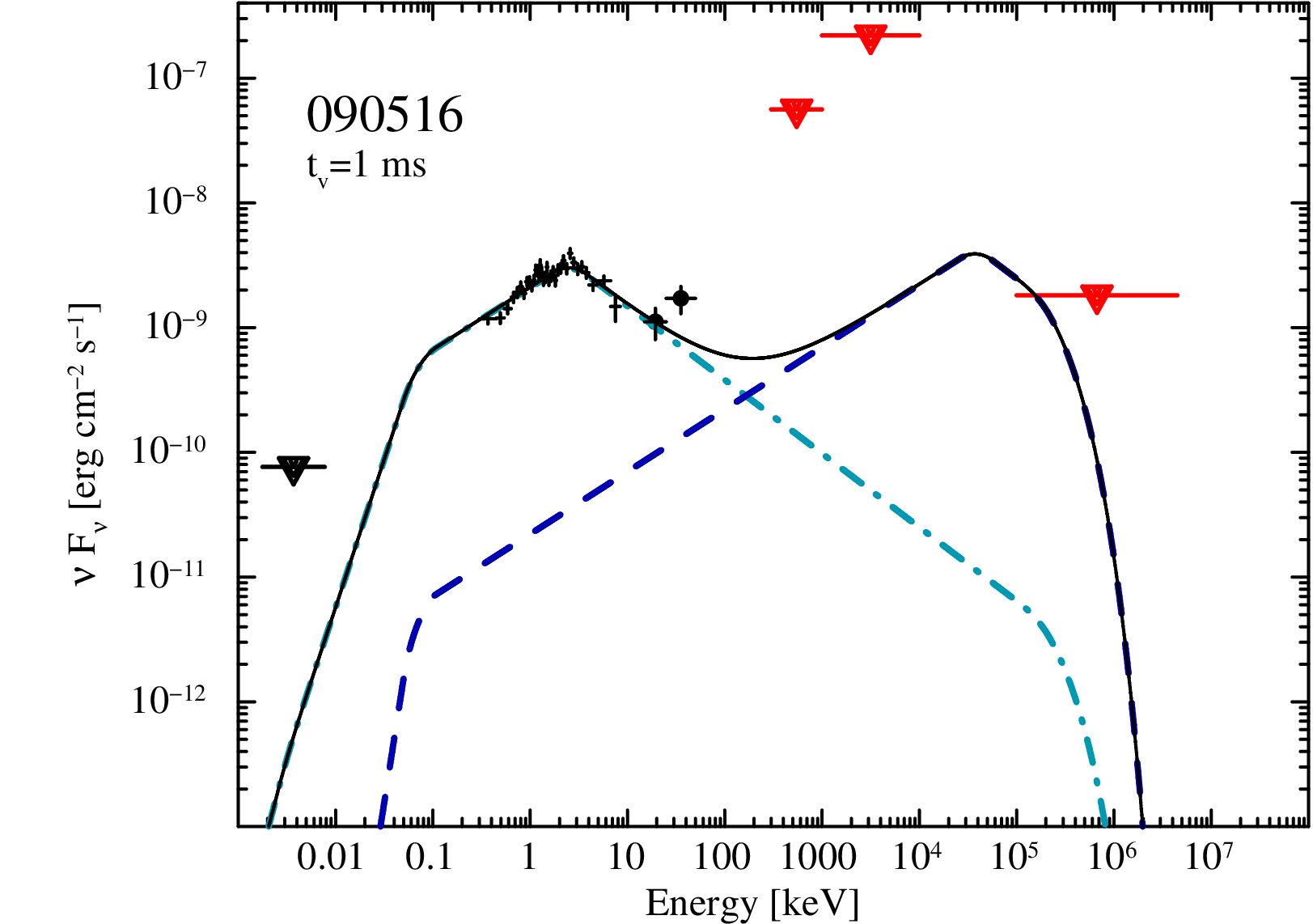}\\
\vspace{0.5cm}
\includegraphics[angle=270,scale=0.31]{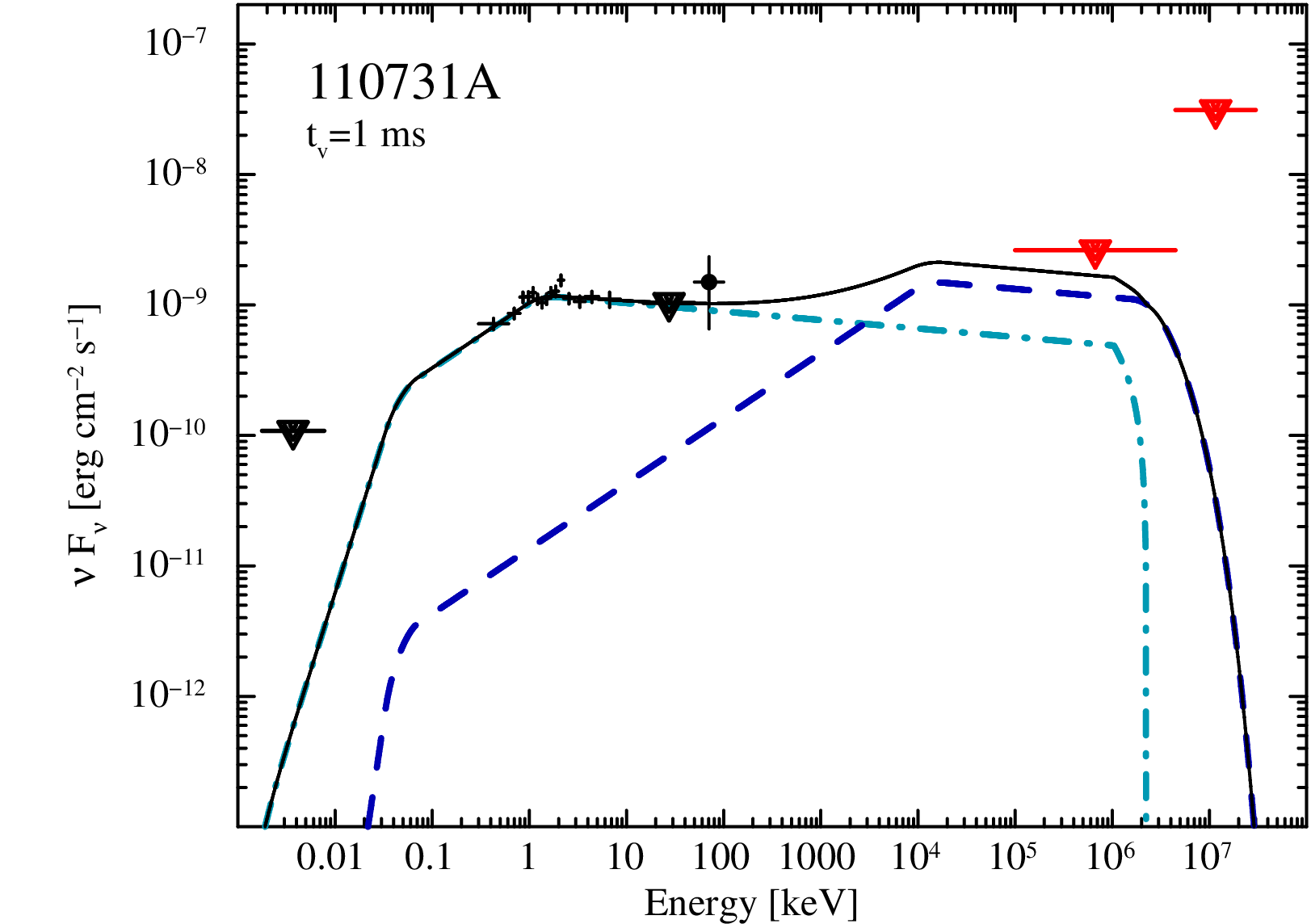}
 \caption{Broadband spectra of X-ray flares comparing \swift~(black) and 
 \fermi~(red) data with the standard IS model.
The best fit model for $t_v$=1~ms is shown by the solid line. 
The synchrotron component is shown by the dot-dashed line, 
the IC component by the dashed line.
The LAT upper limits were calculated in two energy bands, 
100 MeV - 3 GeV and 3 GeV - 30 GeV, 
using the procedure described in Section.~\ref{lat}, and converted into
energy flux units using the best fit spectral model. In the case of GRB~081203A and 090516,
upper limits above 3 GeV are out of scale, and are not shown.}
\vspace{0.5cm}
\label{bestfitms}
\end{figure*}

Now we turn to the detailed description of spectral modelling of the bursts with known redshift. 
The lack of a high-energy counterpart removes a relevant constraint to the model.
In order to limit the number of free parameters we performed the spectral fits by fixing the 
variability timescales to three different values, $t_v$=$\Delta t / (1+z)$, that is the flare duration, 
$t_v$=0.1~s, and $t_v$=1~ms, similar to the variability timescale of the prompt emission. 
The former choice did not provide any acceptable fit unless we introduced a substantial extinction
($A_V>>1$). This is because, within the internal shock scenario, a long variability timescale 
would predict a low self-absorption frequency and therefore an optical flux much larger than the measured values. 
This can be consistent
with the observations only if a significant amount of dust suppresses the optical
emission.
As no afterglow in our sample shows evidence of such a
feature, the absorber should be closer to the central source, where dust cannot survive the strong photon
flux produced by the GRB. Therefore, within the internal shock scenario, 
we found that the condition $t_v$$\approx$$\Delta t$ cannot reproduce the observed data.
This result can be understood by examining in more detail how the constraints on the optical
and gamma-ray fluxes relate to the variability time scale $t_v$. 
Equation~(\ref{casoa}) allows us to derive an upper limit $t_v \lesssim 0.2$\,
$(h \nu_{m}/1~{\rm keV})^{-7/4}$~s,
that applies if the X-ray flare emission is dominated by the synchrotron component. 
If 
the flare duration represents its typical variability timescale, 
the Thomson opacity constraint
requires $h\nu_{m} \ll 1$\,kev. Thus, assuming that the X-ray emission is
dominated by the synchrotron tail at $\nu\gg\nu_m$, the flux at
lower energies would increase with decreasing energy down to
$\nu_ {sa}$. For large values of $t_v$, $\nu_{sa}$
(Equation~\ref{nusa}) is much lower than the X-ray range, and the flux
predicted in the optical band would violate the observed upper
limits. 
If  no substantial extinction is present, 
$\nu_{sa}$ has to be large in order not to overcome the optical limits.  
This condition implies again small values of $t_v \ll \Delta t$. 
The assumption that the flare duration reflects the typical timescale of the relativistic outflow
holds if the X-ray flare emission is mostly produced by IC processes.
As discussed in the previous section, this is in constrast with the \fermi~observations.

In Table~\ref{tab:fit} we report the best fit results for the shorter timescales, 
$t_v=0.1$\,s and $t_v=1$\,ms. 
 Based on our dataset we cannot discriminate between these two values, 
and both fits represent an acceptable description of the data.
The broadband spectra and best-fit models are shown in Fig.~\ref{bestfit}, 
and in Fig.~\ref{bestfitms}, respectively. 
The cases of GRB~080906 and GRB~110731A are the least constrained, as only a
simple power-law component is present in the X-ray spectrum (see
Tab.~1).
In the cases of GRB~080928 and GRB~090516 the peak of the synchrotron component
lies in the XRT band, allowing for better constraints on the parameters.
Above the peak, the photon indices are $\Gamma$=$p/2 + 1$$\sim$2.1 and 2.6,
respectively, consistent with the phenomenogical fits of Tab.~1.
The derived Lorentz factors range between 50 and 120 for $t_v$=0.1\,s, corresponding
to a radius $R$$\approx$\e{13}-\e{14}\,cm, in agreement with the 
late internal shock scenario \citep{fan05}.
Larger Lorentz factors are required for $t_v$=1\,ms, ranging
from 200 to 1000, which correspond to 
$R$$\approx$2\ee{12}-6\ee{13}\,cm.
As a comparison, a Lorentz factor $\Gamma$$\sim$300-550 
was derived for the prompt gamma-ray emission phase
of GRB~110731A \citep{110731a}.
Independent on the value of $t_v$, we found 
0.1\,$\lesssim$\,$\epsilon_B/\epsilon_e$\,$\lesssim$\,1,
and therefore no bright SSC component is expected at high energies.
The typical \fermi~GBM upper limit does not exclude the presence of a 
MeV counterpart 100 times brighter than the observed X-ray flare. 
However, the emergence of such a component would also affect the BAT spectrum,
which, thanks to the sensitivity of the \swift/BAT, provides a
tight constraint on the models. 
   
GRB~081203A is the only case characterized by a turn up in the BAT spectrum,
modeled as the rise of the SSC emission. 
Contrary to the other cases, a small value of the magnetic field, 
$\epsilon_B$$\approx$\e{-2}, 
and a larger value of $Y\gg1$ are needed to account for the observed hard X-ray emission.
As $h \nu_m$$\approx$1~keV, the opacity requirement (Eq.~\ref{casob}) 
yields $t_{v,{\rm max}}$\,$\propto$\,0.08\,$Y^{-4/(p+4)}$\,s, and favors the 
shortest variability timescales.
In fact, for this flare no acceptable fit was found for $t_v$=0.1\,s, and
a variability timescale as short as 1~ms provided a better description. 
This result holds if the emission above 15~keV is associated to the flare. A different possibility
is that the underlying afterglow, subdominant in the soft X-ray band, becomes visible in the
BAT energy range. In the latter hypothesis, no bright SSC flare component is required, and the
properties of this X-ray flare are analoguous to rest of the sample. 

\newpage
\section{Conclusion}\label{sec:end}
We presented a systematic search for high-energy emission
associated with X-ray flares. 
We found that in general X-ray flares are not 
accompanied by a bright counterpart in the MeV-GeV energy range.
By assuming a flat power-law energy spectrum, we derived typical 
upper limits of \e{-7}\,erg\,cm$^{-2}$\,s 
in the 1~MeV-10 MeV energy band, 
and \e{-8}\,erg\,cm$^{-2}$\,s in the 100~MeV-10~GeV band.
Interestingly, no bright optical counterpart is 
observed during the periods of flaring activity. 
The lack of optical and gamma-ray detections disfavors
IC processes as the main radiation mechanism producing
the observed X-ray emission.
 
For bursts with a measured redshift, 
we carried out a more detailed analysis in the
context of the internal shock model.
This model is a good fit for all the
flares and is consistent with the canonical scenario
where the relativistic ($\Gamma>50$) outflow first undergoes internal shocks 
at a radius $\approx$\e{13}-\e{14}\,cm and
then, at larger radii, external shocks. 
X-ray flares carry a substantial fraction of the radiated 
energy, ranging from $\approx 10^{51}$ erg to $\approx 10^{53}$ erg (isotropic equivalent),
that is from 6\% to 100\% of the energy observed during the prompt 
gamma-ray phase. 

Particularly compelling are the implications on the variability
timescale, which in all cases was significantly shorter than
the flare duration. %
The broadband data are not consistent with the simplest scenario,
in which the flare is caused by a ``single event'',
produced by the interaction of two shells 
colliding at a radius $\approx 2 \Gamma^2 c \Delta t $.
The flare light curve reflects instead the profile
of the relativistic outflow, modulated by the central engine 
on a timescale $t_v\lesssim 0.1$\,s.
Longer time scales for variability
would be allowed if the synchrotron emissio peaked
significantly below the X-ray band, but this condition 
is not consistent with the optical and gamma-ray upper limits.
Within the internal shock model considered here, the duration of the 
X-ray flare is mainly set by the prolonged activity of the inner engine.
This is a strong requirement for GRB central engines, as the observed 
durations of X-ray flares are often comparable or even exceed the durations
of the prompt gamma-ray phases.

\section*{Acknowledgements}
The \textit{Fermi} LAT Collaboration acknowledges generous ongoing support
from a number of agencies and institutes that have supported both the
development and the operation of the LAT as well as scientific data analysis.
These include the National Aeronautics and Space Administration and the
Department of Energy in the United States, the Commissariat \`a l'Energie Atomique
and the Centre National de la Recherche Scientifique / Institut National de Physique
Nucl\'eaire et de Physique des Particules in France, the Agenzia Spaziale Italiana
and the Istituto Nazionale di Fisica Nucleare in Italy, the Ministry of Education,
Culture, Sports, Science and Technology (MEXT), High Energy Accelerator Research
Organization (KEK) and Japan Aerospace Exploration Agency (JAXA) in Japan, and
the K.~A.~Wallenberg Foundation, the Swedish Research Council and the
Swedish National Space Board in Sweden.

Additional support for science analysis during the operations phase is gratefully
acknowledged from the Istituto Nazionale di Astrofisica in Italy and the Centre National d'\'Etudes Spatiales in France.

\nocite{*}
\bibliographystyle{aa}

\begin{thebibliography}{77}
\expandafter\ifx\csname natexlab\endcsname\relax\def\natexlab#1{#1}\fi

\bibitem[{{Abdo} {et~al.}(2011){Abdo}, {Ackermann}, {Ajello}, {Baldini},
  {Ballet}, {Barbiellini}, {Baring}, {Bastieri}, {Bechtol}, {Bellazzini},
  {Berenji}, {Bhat}, {Bissaldi}, {Blandford}, {Bonamente}, {Bonnell},
  {Borgland}, {Bouvier}, {Bregeon}, {Brigida}, {Bruel}, {Buehler}, {Buson},
  {Caliandro}, {Cameron}, {Caraveo}, {Casandjian}, {Cecchi}, {Charles},
  {Chekhtman}, {Chiang}, {Ciprini}, {Claus}, {Connaughton}, {Conrad}, {Cutini},
  {de Angelis}, {de Palma}, {Dermer}, {Silva}, {Drell}, {Dubois}, {Favuzzi},
  {Fukazawa}, {Fusco}, {Gargano}, {Gehrels}, {Germani}, {Giglietto}, {Giommi},
  {Giordano}, {Giroletti}, {Glanzman}, {Godfrey}, {Granot}, {Grenier},
  {Guiriec}, {Hadasch}, {Hanabata}, {Hughes}, {J{\'o}hannesson}, {Johnson},
  {Kamae}, {Katagiri}, {Kataoka}, {Kerr}, {Kn{\"o}dlseder}, {Kuss}, {Lande},
  {Latronico}, {Lee}, {Longo}, {Loparco}, {Lott}, {Lubrano}, {Mazziotta},
  {McEnery}, {M{\'e}sz{\'a}ros}, {Michelson}, {Mizuno}, {Moiseev}, {Monzani},
  {Morselli}, {Moskalenko}, {Murgia}, {Nakamori}, {Naumann-Godo}, {Nolan},
  {Norris}, {Nuss}, {Ohsugi}, {Okumura}, {Omodei}, {Orlando}, {Paciesas},
  {Pelassa}, {Pesce-Rollins}, {Pierbattista}, {Piron}, {Porter}, {Racusin},
  {Rain{\`o}}, {Razzano}, {Razzaque}, {Reimer}, {Reimer}, {Reyes}, {Roth},
  {Sadrozinski}, {Sgr{\`o}}, {Siskind}, {Smith}, {Sonbas}, {Spandre},
  {Spinelli}, {Stamatikos}, {Strickman}, {Takahashi}, {Tanaka}, {Tanaka},
  {Thayer}, {Thayer}, {Torres}, {Tosti}, {Troja}, {Uehara}, {Usher},
  {Vandenbroucke}, {Vasileiou}, {Vianello}, {Vilchez}, {Vitale}, {von Kienlin},
  {Waite}, {Wang}, {Winer}, {Wood}, {Yamazaki}, {Yang}, {Ziegler}, \&
  {Piro}}]{100728a}
{Abdo}, A.~A., {Ackermann}, M., {Ajello}, M., {et~al.} 2011, \apjl, 734, L27+

\bibitem[{{Abdo} {et~al.}(2009){Abdo}, {Ackermann}, {Asano}, {Atwood},
  {Axelsson}, {Baldini}, {Ballet}, {Band}, {Barbiellini}, {Bastieri},
  {Bechtol}, {Bellazzini}, {Berenji}, {Bhat}, {Bissaldi}, {Bloom}, {Bonamente},
  {Borgland}, {Bouvier}, {Bregeon}, {Brez}, {Briggs}, {Brigida}, {Bruel},
  {Burnett}, {Caliandro}, {Cameron}, {Caraveo}, {Casandjian}, {Cecchi},
  {Chaplin}, {Chekhtman}, {Cheung}, {Chiang}, {Ciprini}, {Claus},
  {Cohen-Tanugi}, {Cominsky}, {Connaughton}, {Conrad}, {Cutini}, {Dermer}, {de
  Angelis}, {de Palma}, {Digel}, {Silva}, {Drell}, {Dubois}, {Dumora},
  {Farnier}, {Favuzzi}, {Focke}, {Frailis}, {Fukazawa}, {Fusco}, {Gargano},
  {Gasparrini}, {Gehrels}, {Germani}, {Gibby}, {Giebels}, {Giglietto},
  {Giordano}, {Glanzman}, {Godfrey}, {Goldstein}, {Granot}, {Grenier},
  {Grondin}, {Grove}, {Guillemot}, {Guiriec}, {Hanabata}, {Harding},
  {Hayashida}, {Hays}, {Hughes}, {J{\'o}hannesson}, {Johnson}, {Johnson},
  {Kamae}, {Katagiri}, {Kataoka}, {Kawai}, {Kerr}, {Kn{\"o}dlseder},
  {Kocevski}, {Komin}, {Kouveliotou}, {Kuehn}, {Kuss}, {Latronico}, {Longo},
  {Loparco}, {Lott}, {Lovellette}, {Lubrano}, {Makeev}, {Mazziotta}, {McBreen},
  {McEnery}, {McGlynn}, {Meegan}, {Meurer}, {Michelson}, {Mitthumsiri},
  {Mizuno}, {Monte}, {Monzani}, {Moretti}, {Morselli}, {Moskalenko}, {Murgia},
  {Nakamori}, {Nolan}, {Norris}, {Nuss}, {Ohno}, {Ohsugi}, {Omodei}, {Orlando},
  {Ormes}, {Ozaki}, {Paciesas}, {Paneque}, {Panetta}, {Parent}, {Pelassa},
  {Pepe}, {Pesce-Rollins}, {Piron}, {Porter}, {Preece}, {Rain{\`o}}, {Rando},
  {Razzano}, {Razzaque}, {Reimer}, {Reposeur}, {Ritz}, {Rochester},
  {Rodriguez}, {Roth}, {Ryde}, {Sadrozinski}, {Sanchez}, {Sander}, {Saz
  Parkinson}, {Scargle}, {Sgr{\`o}}, {Siskind}, {Smith}, {Smith}, {Spandre},
  {Spinelli}, {Stamatikos}, {Strickman}, {Suson}, {Tajima}, {Takahashi},
  {Tanaka}, {Thayer}, {Thayer}, {Tibaldo}, {Torres}, {Tosti}, {Tramacere},
  {Uchiyama}, {Usher}, {van der Horst}, {Vasileiou}, {Vilchez}, {Vitale}, {von
  Kienlin}, {Waite}, {Wang}, {Wilson-Hodge}, {Winer}, {Wood}, {Ylinen}, \&
  {Ziegler}}]{080825c}
{Abdo}, A.~A., {Ackermann}, M., {Asano}, K., {et~al.} 2009, \apj, 707, 580

\bibitem[{{Ackermann} {et~al.}(2012){Ackermann}, {Ajello}, {Albert},
  {Allafort}, {Atwood}, {Axelsson}, {Baldini}, {Ballet}, {Barbiellini},
  {Bastieri}, {Bechtol}, {Bellazzini}, {Bissaldi}, {Blandford}, {Bloom},
  {Bogart}, {Bonamente}, {Borgland}, {Bottacini}, {Bouvier}, {Brandt},
  {Bregeon}, {Brigida}, {Bruel}, {Buehler}, {Burnett}, {Buson}, {Caliandro},
  {Cameron}, {Caraveo}, {Casandjian}, {Cavazzuti}, {Cecchi}, {{\c C}elik},
  {Charles}, {Chaves}, {Chekhtman}, {Cheung}, {Chiang}, {Ciprini}, {Claus},
  {Cohen-Tanugi}, {Conrad}, {Corbet}, {Cutini}, {D'Ammando}, {Davis}, {de
  Angelis}, {DeKlotz}, {de Palma}, {Dermer}, {Digel}, {Silva}, {Drell},
  {Drlica-Wagner}, {Dubois}, {Favuzzi}, {Fegan}, {Ferrara}, {Focke}, {Fortin},
  {Fukazawa}, {Funk}, {Fusco}, {Gargano}, {Gasparrini}, {Gehrels}, {Giebels},
  {Giglietto}, {Giordano}, {Giroletti}, {Glanzman}, {Godfrey}, {Grenier},
  {Grove}, {Guiriec}, {Hadasch}, {Hayashida}, {Hays}, {Horan}, {Hou}, {Hughes},
  {Jackson}, {Jogler}, {J{\'o}hannesson}, {Johnson}, {Johnson}, {Johnson},
  {Kamae}, {Katagiri}, {Kataoka}, {Kerr}, {Kn{\"o}dlseder}, {Kuss}, {Lande},
  {Larsson}, {Latronico}, {Lavalley}, {Lemoine-Goumard}, {Longo}, {Loparco},
  {Lott}, {Lovellette}, {Lubrano}, {Mazziotta}, {McConville}, {McEnery},
  {Mehault}, {Michelson}, {Mitthumsiri}, {Mizuno}, {Moiseev}, {Monte},
  {Monzani}, {Morselli}, {Moskalenko}, {Murgia}, {Naumann-Godo}, {Nemmen},
  {Nishino}, {Norris}, {Nuss}, {Ohno}, {Ohsugi}, {Okumura}, {Omodei},
  {Orienti}, {Orlando}, {Ormes}, {Paneque}, {Panetta}, {Perkins},
  {Pesce-Rollins}, {Pierbattista}, {Piron}, {Pivato}, {Porter}, {Racusin},
  {Rain{\`o}}, {Rando}, {Razzano}, {Razzaque}, {Reimer}, {Reimer}, {Reposeur},
  {Reyes}, {Ritz}, {Rochester}, {Romoli}, {Roth}, {Sadrozinski}, {Sanchez},
  {Saz Parkinson}, {Sbarra}, {Scargle}, {Sgr{\`o}}, {Siegal-Gaskins},
  {Siskind}, {Spandre}, {Spinelli}, {Stephens}, {Suson}, {Tajima}, {Takahashi},
  {Tanaka}, {Thayer}, {Thayer}, {Thompson}, {Tibaldo}, {Tinivella}, {Tosti},
  {Troja}, {Usher}, {Vandenbroucke}, {Van Klaveren}, {Vasileiou}, {Vianello},
  {Vitale}, {Waite}, {Wallace}, {Winer}, {Wood}, {Wood}, {Wood}, {Yang}, \&
  {Zimmer}}]{pass7}
{Ackermann}, M., {Ajello}, M., {Albert}, A., {et~al.} 2012, \apjs, 203, 4

\bibitem[{{Ackermann} {et~al.}(2013){Ackermann}, {Ajello}, {Asano}, {Baldini},
  {Barbiellini}, {Baring}, {Bastieri}, {Bellazzini}, {Blandford}, {Bonamente},
  {Borgland}, {Bottacini}, {Bregeon}, {Brigida}, {Bruel}, {Buehler}, {Buson},
  {Caliandro}, {Cameron}, {Caraveo}, {Cecchi}, {Charles}, {Chaves},
  {Chekhtman}, {Chiang}, {Ciprini}, {Claus}, {Cohen-Tanugi}, {Conrad},
  {Cutini}, {D'Ammando}, {de Angelis}, {de Palma}, {Dermer}, {Silva}, {Drell},
  {Drlica-Wagner}, {Favuzzi}, {Fegan}, {Focke}, {Franckowiak}, {Fukazawa},
  {Fusco}, {Gargano}, {Gasparrini}, {Gehrels}, {Giglietto}, {Giordano},
  {Giroletti}, {Glanzman}, {Godfrey}, {Granot}, {Greiner}, {Grenier}, {Grove},
  {Guiriec}, {Hadasch}, {Hanabata}, {Hayashida}, {Hays}, {Hughes}, {Jackson},
  {Jogler}, {J{\'o}hannesson}, {Johnson}, {Kn{\"o}dlseder}, {Kocevski}, {Kuss},
  {Lande}, {Larsson}, {Latronico}, {Longo}, {Loparco}, {Lovellette}, {Lubrano},
  {Mazziotta}, {McEnery}, {Mehault}, {M{\'e}sz{\'a}ros}, {Michelson},
  {Mitthumsiri}, {Mizuno}, {Monte}, {Monzani}, {Moretti}, {Morselli},
  {Moskalenko}, {Murgia}, {Naumann-Godo}, {Norris}, {Nuss}, {Nymark}, {Ohno},
  {Ohsugi}, {Omodei}, {Orienti}, {Orlando}, {Paneque}, {Perkins},
  {Pesce-Rollins}, {Piron}, {Pivato}, {Racusin}, {Rain{\`o}}, {Rando},
  {Razzano}, {Razzaque}, {Reimer}, {Reimer}, {Romoli}, {Roth}, {Ryde},
  {Sanchez}, {Sgr{\`o}}, {Siskind}, {Sonbas}, {Spinelli}, {Stamatikos},
  {Takahashi}, {Tanaka}, {Thayer}, {Thayer}, {Tibaldo}, {Tinivella}, {Tosti},
  {Troja}, {Usher}, {Vandenbroucke}, {Vasileiou}, {Vianello}, {Vitale},
  {Waite}, {Winer}, {Wood}, {Yang}, {Gruber}, {Bhat}, {Bissaldi}, {Briggs},
  {Burgess}, {Connaughton}, {Foley}, {Kippen}, {Kouveliotou}, {McBreen},
  {McGlynn}, {Paciesas}, {Pelassa}, {Preece}, {Rau}, {van der Horst}, {von
  Kienlin}, {Kann}, {Filgas}, {Klose}, {Kr{\"u}hler}, {Fukui}, {Sako},
  {Tristram}, {Oates}, {Ukwatta}, \& {Littlejohns}}]{110731a}
{Ackermann}, M., {Ajello}, M., {Asano}, K., {et~al.} 2013, \apj, 763, 71

\bibitem[{{Ackermann et al.}(2013)}]{latcat}
{Ackermann et al.} 2013, \apjs, 209, 11

\bibitem[{{Arnaud} {et~al.}(2011){Arnaud}, {Smith}, \&
  {Siemiginowska}}]{arnaud11}
{Arnaud}, K., {Smith}, R., \& {Siemiginowska}, A. 2011, {Handbook of X-ray
  Astronomy}, ed. R.~{Ellis}, J.~{Huchra}, S.~{Kahn}, G.~{Rieke}, \& P.~B.
  {Stetson}

\bibitem[{{Arnaud}(1996)}]{arnaud96}
{Arnaud}, K.~A. 1996, in Astronomical Society of the Pacific Conference Series,
  Vol. 101, Astronomical Data Analysis Software and Systems V, ed.
  {G.~H.~Jacoby \& J.~Barnes}, 17--+

\bibitem[{{Atwood} {et~al.}(2009){Atwood}, {Abdo}, {Ackermann}, {Althouse},
  {Anderson}, {Axelsson}, {Baldini}, {Ballet}, {Band}, {Barbiellini}, \&
  et~al.}]{atwood09}
{Atwood}, W.~B., {Abdo}, A.~A., {Ackermann}, M., {et~al.} 2009, \apj, 697, 1071

\bibitem[{{Band} {et~al.}(1993){Band}, {Matteson}, {Ford}, {Schaefer},
  {Palmer}, {Teegarden}, {Cline}, {Briggs}, {Paciesas}, {Pendleton}, {Fishman},
  {Kouveliotou}, {Meegan}, {Wilson}, \& {Lestrade}}]{band93}
{Band}, D., {Matteson}, J., {Ford}, L., {et~al.} 1993, \apj, 413, 281

\bibitem[{{Baring} \& {Harding}(1997)}]{baring97}
{Baring}, M.~G. \& {Harding}, A.~K. 1997, \apj, 491, 663

\bibitem[{{Barthelmy} {et~al.}(2005{\natexlab{a}}){Barthelmy}, {Barbier},
  {Cummings}, {Fenimore}, {Gehrels}, {Hullinger}, {Krimm}, {Markwardt},
  {Palmer}, {Parsons}, {Sato}, {Suzuki}, {Takahashi}, {Tashiro}, \&
  {Tueller}}]{bat05}
{Barthelmy}, S.~D., {Barbier}, L.~M., {Cummings}, J.~R., {et~al.}
  2005{\natexlab{a}}, \ssr, 120, 143

\bibitem[{{Barthelmy} {et~al.}(2005{\natexlab{b}}){Barthelmy}, {Chincarini},
  {Burrows}, {Gehrels}, {Covino}, {Moretti}, {Romano}, {O'Brien}, {Sarazin},
  {Kouveliotou}, {Goad}, {Vaughan}, {Tagliaferri}, {Zhang}, {Antonelli},
  {Campana}, {Cummings}, {D'Avanzo}, {Davies}, {Giommi}, {Grupe}, {Kaneko},
  {Kennea}, {King}, {Kobayashi}, {Melandri}, {Meszaros}, {Nousek}, {Patel},
  {Sakamoto}, \& {Wijers}}]{barthelmy05}
{Barthelmy}, S.~D., {Chincarini}, G., {Burrows}, D.~N., {et~al.}
  2005{\natexlab{b}}, \nat, 438, 994

\bibitem[{{Beloborodov} {et~al.}(2011){Beloborodov}, {Daigne}, {Mochkovitch},
  \& {Uhm}}]{belo11}
{Beloborodov}, A.~M., {Daigne}, F., {Mochkovitch}, R., \& {Uhm}, Z.~L. 2011,
  \mnras, 410, 2422

\bibitem[{{Blumenthal} \& {Gould}(1970)}]{bg70}
{Blumenthal}, G.~R. \& {Gould}, R.~J. 1970, Reviews of Modern Physics, 42, 237

\bibitem[{{Breeveld} {et~al.}(2010){Breeveld}, {Curran}, {Hoversten}, {Koch},
  {Landsman}, {Marshall}, {Page}, {Poole}, {Roming}, {Smith}, {Still},
  {Yershov}, {Blustin}, {Brown}, {Gronwall}, {Holland}, {Kuin}, {McGowan},
  {Rosen}, {Boyd}, {Broos}, {Carter}, {Chester}, {Hancock}, {Huckle}, {Immler},
  {Ivanushkina}, {Kennedy}, {Mason}, {Morgan}, {Oates}, {de Pasquale},
  {Schady}, {Siegel}, \& {vanden Berk}}]{breeveld10}
{Breeveld}, A.~A., {Curran}, P.~A., {Hoversten}, E.~A., {et~al.} 2010, \mnras,
  406, 1687

\bibitem[{{Breeveld} {et~al.}(2011){Breeveld}, {Landsman}, {Holland}, {Roming},
  {Kuin}, \& {Page}}]{breeveld11}
{Breeveld}, A.~A., {Landsman}, W., {Holland}, S.~T., {et~al.} 2011, in American
  Institute of Physics Conference Series, Vol. 1358, American Institute of
  Physics Conference Series, ed. J.~E. {McEnery}, J.~L. {Racusin}, \&
  N.~{Gehrels}, 373--376

\bibitem[{{Burrows} {et~al.}(2005{\natexlab{a}}){Burrows}, {Hill}, {Nousek},
  {Kennea}, {Wells}, {Osborne}, {Abbey}, {Beardmore}, {Mukerjee}, {Short},
  {Chincarini}, {Campana}, {Citterio}, {Moretti}, {Pagani}, {Tagliaferri},
  {Giommi}, {Capalbi}, {Tamburelli}, {Angelini}, {Cusumano}, {Br{\"a}uninger},
  {Burkert}, \& {Hartner}}]{xrt05}
{Burrows}, D.~N., {Hill}, J.~E., {Nousek}, J.~A., {et~al.} 2005{\natexlab{a}},
  \ssr, 120, 165

\bibitem[{{Burrows} {et~al.}(2005{\natexlab{b}}){Burrows}, {Romano}, {Falcone},
  {Kobayashi}, {Zhang}, {Moretti}, {O'Brien}, {Goad}, {Campana}, {Page},
  {Angelini}, {Barthelmy}, {Beardmore}, {Capalbi}, {Chincarini}, {Cummings},
  {Cusumano}, {Fox}, {Giommi}, {Hill}, {Kennea}, {Krimm}, {Mangano},
  {Marshall}, {M{\'e}sz{\'a}ros}, {Morris}, {Nousek}, {Osborne}, {Pagani},
  {Perri}, {Tagliaferri}, {Wells}, {Woosley}, \& {Gehrels}}]{burrows05}
{Burrows}, D.~N., {Romano}, P., {Falcone}, A., {et~al.} 2005{\natexlab{b}},
  Science, 309, 1833

\bibitem[{{Chincarini} {et~al.}(2007){Chincarini}, {Moretti}, {Romano},
  {Falcone}, {Morris}, {Racusin}, {Campana}, {Covino}, {Guidorzi},
  {Tagliaferri}, {Burrows}, {Pagani}, {Stroh}, {Grupe}, {Capalbi}, {Cusumano},
  {Gehrels}, {Giommi}, {La Parola}, {Mangano}, {Mineo}, {Nousek}, {O'Brien},
  {Page}, {Perri}, {Troja}, {Willingale}, \& {Zhang}}]{chinca07}
{Chincarini}, G., {Moretti}, A., {Romano}, P., {et~al.} 2007, \apj, 671, 1903

\bibitem[{{Curran} {et~al.}(2008){Curran}, {Starling}, {O'Brien}, {Godet}, {van
  der Horst}, \& {Wijers}}]{curran08}
{Curran}, P.~A., {Starling}, R.~L.~C., {O'Brien}, P.~T., {et~al.} 2008, \aap,
  487, 533

\bibitem[{{Evans} {et~al.}(2009){Evans}, {Beardmore}, {Page}, {Osborne},
  {O'Brien}, {Willingale}, {Starling}, {Burrows}, {Godet}, {Vetere}, {Racusin},
  {Goad}, {Wiersema}, {Angelini}, {Capalbi}, {Chincarini}, {Gehrels}, {Kennea},
  {Margutti}, {Morris}, {Mountford}, {Pagani}, {Perri}, {Romano}, \&
  {Tanvir}}]{evans09}
{Evans}, P.~A., {Beardmore}, A.~P., {Page}, K.~L., {et~al.} 2009, \mnras, 397,
  1177

\bibitem[{{Evans} {et~al.}(2007){Evans}, {Beardmore}, {Page}, {Tyler},
  {Osborne}, {Goad}, {O'Brien}, {Vetere}, {Racusin}, {Morris}, {Burrows},
  {Capalbi}, {Perri}, {Gehrels}, \& {Romano}}]{evans07}
{Evans}, P.~A., {Beardmore}, A.~P., {Page}, K.~L., {et~al.} 2007, \aap, 469,
  379

\bibitem[{{Falcone} {et~al.}(2006){Falcone}, {Burrows}, {Lazzati}, {Campana},
  {Kobayashi}, {Zhang}, {M{\'e}sz{\'a}ros}, {Page}, {Kennea}, {Romano},
  {Pagani}, {Angelini}, {Beardmore}, {Capalbi}, {Chincarini}, {Cusumano},
  {Giommi}, {Goad}, {Godet}, {Grupe}, {Hill}, {La Parola}, {Mangano},
  {Moretti}, {Nousek}, {O'Brien}, {Osborne}, {Perri}, {Tagliaferri}, {Wells},
  \& {Gehrels}}]{falcone06}
{Falcone}, A.~D., {Burrows}, D.~N., {Lazzati}, D., {et~al.} 2006, \apj, 641,
  1010

\bibitem[{{Falcone} {et~al.}(2007){Falcone}, {Morris}, {Racusin}, {Chincarini},
  {Moretti}, {Romano}, {Burrows}, {Pagani}, {Stroh}, {Grupe}, {Campana},
  {Covino}, {Tagliaferri}, {Willingale}, \& {Gehrels}}]{falcone07}
{Falcone}, A.~D., {Morris}, D., {Racusin}, J., {et~al.} 2007, \apj, 671, 1921

\bibitem[{{Fan} {et~al.}(2008){Fan}, {Piran}, {Narayan}, \& {Wei}}]{fan08}
{Fan}, Y.-Z., {Piran}, T., {Narayan}, R., \& {Wei}, D.-M. 2008, \mnras, 384,
  1483

\bibitem[{{Fan} \& {Wei}(2005)}]{fan05}
{Fan}, Y.~Z. \& {Wei}, D.~M. 2005, \mnras, 364, L42

\bibitem[{{Fryer} \& {M{\'e}sz{\'a}ros}(2003)}]{fryer03}
{Fryer}, C.~L. \& {M{\'e}sz{\'a}ros}, P. 2003, \apjl, 588, L25

\bibitem[{{Galli} \& {Piro}(2006)}]{galli06}
{Galli}, A. \& {Piro}, L. 2006, \aap, 455, 413

\bibitem[{{Galli} \& {Piro}(2007)}]{galli07}
{Galli}, A. \& {Piro}, L. 2007, \aap, 475, 421

\bibitem[{{Gehrels} {et~al.}(2004){Gehrels}, {Chincarini}, {Giommi}, {Mason},
  {Nousek}, {Wells}, {White}, {Barthelmy}, {Burrows}, {Cominsky}, {Hurley},
  {Marshall}, {M{\'e}sz{\'a}ros}, {Roming}, {Angelini}, {Barbier}, {Belloni},
  {Campana}, {Caraveo}, {Chester}, {Citterio}, {Cline}, {Cropper}, {Cummings},
  {Dean}, {Feigelson}, {Fenimore}, {Frail}, {Fruchter}, {Garmire}, {Gendreau},
  {Ghisellini}, {Greiner}, {Hill}, {Hunsberger}, {Krimm}, {Kulkarni}, {Kumar},
  {Lebrun}, {Lloyd-Ronning}, {Markwardt}, {Mattson}, {Mushotzky}, {Norris},
  {Osborne}, {Paczynski}, {Palmer}, {Park}, {Parsons}, {Paul}, {Rees},
  {Reynolds}, {Rhoads}, {Sasseen}, {Schaefer}, {Short}, {Smale}, {Smith},
  {Stella}, {Tagliaferri}, {Takahashi}, {Tashiro}, {Townsley}, {Tueller},
  {Turner}, {Vietri}, {Voges}, {Ward}, {Willingale}, {Zerbi}, \&
  {Zhang}}]{swift}
{Gehrels}, N., {Chincarini}, G., {Giommi}, P., {et~al.} 2004, \apj, 611, 1005

\bibitem[{{Giannios}(2006)}]{giannios06}
{Giannios}, D. 2006, \aap, 455, L5

\bibitem[{{Goad} {et~al.}(2007){Goad}, {Page}, {Godet}, {Beardmore}, {Osborne},
  {O'Brien}, {Starling}, {Holland}, {Band}, {Falcone}, {Gehrels}, {Burrows},
  {Nousek}, {Roming}, {Moretti}, \& {Perri}}]{goad07}
{Goad}, M.~R., {Page}, K.~L., {Godet}, O., {et~al.} 2007, \aap, 468, 103

\bibitem[{{Granot} {et~al.}(2000){Granot}, {Piran}, \& {Sari}}]{gps00}
{Granot}, J., {Piran}, T., \& {Sari}, R. 2000, \apjl, 534, L163

\bibitem[{{Guetta} {et~al.}(2007){Guetta}, {Fiore}, {D'Elia}, {Perna},
  {Antonelli}, {Piranomonte}, {Puccetti}, {Stella}, {Angelini}, {Schartel},
  {Campana}, {Chincarini}, {Covino}, {Tagliaferri}, {Malesani}, {Guidorzi},
  {Monfardini}, {Mundell}, {de Le{\'o}n Cruz}, {Castro-Tirado}, {Guzly},
  {Gorosabel}, {Jelinek}, \& {Gomboc}}]{guetta07}
{Guetta}, D., {Fiore}, F., {D'Elia}, V., {et~al.} 2007, \aap, 461, 95

\bibitem[{{Guetta} \& {Granot}(2003)}]{gg03}
{Guetta}, D. \& {Granot}, J. 2003, \apj, 585, 885

\bibitem[{{He} {et~al.}(2012){He}, {Zhang}, {Wang}, {Li}, \&
  {M{\'e}sz{\'a}ros}}]{he11}
{He}, H.-N., {Zhang}, B.-B., {Wang}, X.-Y., {Li}, Z., \& {M{\'e}sz{\'a}ros}, P.
  2012, \apj, 753, 178

\bibitem[{{Ioka} {et~al.}(2005){Ioka}, {Kobayashi}, \& {Zhang}}]{ioka05}
{Ioka}, K., {Kobayashi}, S., \& {Zhang}, B. 2005, \apj, 631, 429

\bibitem[{{Jakobsson} {et~al.}(2004){Jakobsson}, {Hjorth}, {Fynbo}, {Watson},
  {Pedersen}, {Bj{\"o}rnsson}, \& {Gorosabel}}]{jako04}
{Jakobsson}, P., {Hjorth}, J., {Fynbo}, J.~P.~U., {et~al.} 2004, \apjl, 617,
  L21

\bibitem[{{Kalberla} {et~al.}(2005){Kalberla}, {Burton}, {Hartmann}, {Arnal},
  {Bajaja}, {Morras}, \& {P{\"o}ppel}}]{kalberla05}
{Kalberla}, P.~M.~W., {Burton}, W.~B., {Hartmann}, D., {et~al.} 2005, \aap,
  440, 775

\bibitem[{{King} {et~al.}(2005){King}, {O'Brien}, {Goad}, {Osborne}, {Olsson},
  \& {Page}}]{king05}
{King}, A., {O'Brien}, P.~T., {Goad}, M.~R., {et~al.} 2005, \apjl, 630, L113

\bibitem[{{Kobayashi} {et~al.}(2007){Kobayashi}, {Zhang}, {M{\'e}sz{\'a}ros},
  \& {Burrows}}]{koba07}
{Kobayashi}, S., {Zhang}, B., {M{\'e}sz{\'a}ros}, P., \& {Burrows}, D. 2007,
  \apj, 655, 391

\bibitem[{{Krimm} {et~al.}(2007){Krimm}, {Granot}, {Marshall}, {Perri},
  {Barthelmy}, {Burrows}, {Gehrels}, {M{\'e}sz{\'a}ros}, \& {Morris}}]{krimm07}
{Krimm}, H.~A., {Granot}, J., {Marshall}, F.~E., {et~al.} 2007, \apj, 665, 554

\bibitem[{{La Parola} {et~al.}(2006){La Parola}, {Mangano}, {Zhang},
  {Cusumano}, {Mineo}, {Troja}, {Burrows}, {Campana}, {Capalbi}, {Chincarini},
  {Giommi}, {Moretti}, {Perri}, {Romano}, \& {Tagliaferri}}]{laparola06}
{La Parola}, V., {Mangano}, V., {Zhang}, B., {et~al.} 2006, Nuovo Cimento B
  Serie, 121, 1505

\bibitem[{{Lazzati} {et~al.}(2011){Lazzati}, {Blackwell}, {Morsony}, \&
  {Begelman}}]{lazzati11}
{Lazzati}, D., {Blackwell}, C.~H., {Morsony}, B.~J., \& {Begelman}, M.~C. 2011,
  \mnras, 411, L16

\bibitem[{{Lazzati} \& {Perna}(2007)}]{lazzati07}
{Lazzati}, D. \& {Perna}, R. 2007, \mnras, 375, L46

\bibitem[{{Liang} {et~al.}(2006){Liang}, {Zhang}, {O'Brien}, {Willingale},
  {Angelini}, {Burrows}, {Campana}, {Chincarini}, {Falcone}, {Gehrels}, {Goad},
  {Grupe}, {Kobayashi}, {M{\'e}sz{\'a}ros}, {Nousek}, {Osborne}, {Page}, \&
  {Tagliaferri}}]{liang+06}
{Liang}, E.~W., {Zhang}, B., {O'Brien}, P.~T., {et~al.} 2006, \apj, 646, 351

\bibitem[{{Lithwick} \& {Sari}(2001)}]{ls01}
{Lithwick}, Y. \& {Sari}, R. 2001, \apj, 555, 540

\bibitem[{{L{\"u}} \& {Zhang}(2014)}]{lu14}
{L{\"u}}, H.-J. \& {Zhang}, B. 2014, \apj, 785, 74

\bibitem[{{Lyons} {et~al.}(2010){Lyons}, {O'Brien}, {Zhang}, {Willingale},
  {Troja}, \& {Starling}}]{lyons10}
{Lyons}, N., {O'Brien}, P.~T., {Zhang}, B., {et~al.} 2010, \mnras, 402, 705

\bibitem[{{Maxham} \& {Zhang}(2009)}]{maxham09}
{Maxham}, A. \& {Zhang}, B. 2009, \apj, 707, 1623

\bibitem[{{Melandri} {et~al.}(2012){Melandri}, {Sbarufatti}, {D'Avanzo},
  {Salvaterra}, {Campana}, {Covino}, {Vergani}, {Nava}, {Ghisellini},
  {Ghirlanda}, {Fugazza}, {Mangano}, {Capalbi}, \& {Tagliaferri}}]{melandri12}
{Melandri}, A., {Sbarufatti}, B., {D'Avanzo}, P., {et~al.} 2012, \mnras, 421,
  1265

\bibitem[{{Mesler} {et~al.}(2012){Mesler}, {Whalen}, {Lloyd-Ronning}, {Fryer},
  \& {Pihlstr{\"o}m}}]{mesler12}
{Mesler}, R.~A., {Whalen}, D.~J., {Lloyd-Ronning}, N.~M., {Fryer}, C.~L., \&
  {Pihlstr{\"o}m}, Y.~M. 2012, \apj, 757, 117

\bibitem[{{Nakamura}(2010)}]{nakamura10}
{Nakamura}, K. 2010, Journal of Physics G Nuclear Physics, 37, 075021

\bibitem[{{Nakar} {et~al.}(2009){Nakar}, {Ando}, \& {Sari}}]{nakar09}
{Nakar}, E., {Ando}, S., \& {Sari}, R. 2009, \apj, 703, 675

\bibitem[{{Nousek} {et~al.}(2006){Nousek}, {Kouveliotou}, {Grupe}, {Page},
  {Granot}, {Ramirez-Ruiz}, {Patel}, {Burrows}, {Mangano}, {Barthelmy},
  {Beardmore}, {Campana}, {Capalbi}, {Chincarini}, {Cusumano}, {Falcone},
  {Gehrels}, {Giommi}, {Goad}, {Godet}, {Hurkett}, {Kennea}, {Moretti},
  {O'Brien}, {Osborne}, {Romano}, {Tagliaferri}, \& {Wells}}]{nousek06}
{Nousek}, J.~A., {Kouveliotou}, C., {Grupe}, D., {et~al.} 2006, \apj, 642, 389

\bibitem[{{O'Brien} {et~al.}(2006){O'Brien}, {Willingale}, {Osborne}, {Goad},
  {Page}, {Vaughan}, {Rol}, {Beardmore}, {Godet}, {Hurkett}, {Wells}, {Zhang},
  {Kobayashi}, {Burrows}, {Nousek}, {Kennea}, {Falcone}, {Grupe}, {Gehrels},
  {Barthelmy}, {Cannizzo}, {Cummings}, {Hill}, {Krimm}, {Chincarini},
  {Tagliaferri}, {Campana}, {Moretti}, {Giommi}, {Perri}, {Mangano}, \&
  {LaParola}}]{obrien06}
{O'Brien}, P.~T., {Willingale}, R., {Osborne}, J., {et~al.} 2006, \apj, 647,
  1213

\bibitem[{{Page} {et~al.}(2007){Page}, {Willingale}, {Osborne}, {Zhang},
  {Godet}, {Marshall}, {Melandri}, {Norris}, {O'Brien}, {Pal'shin}, {Rol},
  {Romano}, {Starling}, {Schady}, {Yost}, {Barthelmy}, {Beardmore}, {Cusumano},
  {Burrows}, {De Pasquale}, {Ehle}, {Evans}, {Gehrels}, {Goad}, {Golenetskii},
  {Guidorzi}, {Mundell}, {Page}, {Ricker}, {Sakamoto}, {Schaefer},
  {Stamatikos}, {Troja}, {Ulanov}, {Yuan}, \& {Ziaeepour}}]{page07}
{Page}, K.~L., {Willingale}, R., {Osborne}, J.~P., {et~al.} 2007, \apj, 663,
  1125

\bibitem[{{Panaitescu}(2008)}]{pana08}
{Panaitescu}, A. 2008, \mnras, 383, 1143

\bibitem[{{Panaitescu} \& {Kumar}(2000)}]{pk00}
{Panaitescu}, A. \& {Kumar}, P. 2000, \apj, 543, 66

\bibitem[{{Perna} {et~al.}(2006){Perna}, {Armitage}, \& {Zhang}}]{perna06}
{Perna}, R., {Armitage}, P.~J., \& {Zhang}, B. 2006, \apjl, 636, L29

\bibitem[{{Perri} {et~al.}(2007){Perri}, {Guetta}, {Antonelli}, {Cucchiara},
  {Mangano}, {Reeves}, {Angelini}, {Beardmore}, {Boyd}, {Burrows}, {Campana},
  {Capalbi}, {Chincarini}, {Cusumano}, {Giommi}, {Hill}, {Holland}, {La
  Parola}, {Mineo}, {Moretti}, {Nousek}, {Osborne}, {Pagani}, {Romano},
  {Roming}, {Starling}, {Tagliaferri}, {Troja}, {Vetere}, \&
  {Gehrels}}]{perri07}
{Perri}, M., {Guetta}, D., {Antonelli}, L.~A., {et~al.} 2007, \aap, 471, 83

\bibitem[{{Piran}(2004)}]{piran04}
{Piran}, T. 2004, Reviews of Modern Physics, 76, 1143

\bibitem[{{Piro} {et~al.}(1998){Piro}, {Amati}, {Antonelli}, {Butler}, {Costa},
  {Cusumano}, {Feroci}, {Frontera}, {Heise}, {in 't Zand}, {Molendi}, {Muller},
  {Nicastro}, {Orlandini}, {Owens}, {Parmar}, {Soffitta}, \& {Tavani}}]{piro98}
{Piro}, L., {Amati}, L., {Antonelli}, L.~A., {et~al.} 1998, \aap, 331, L41

\bibitem[{{Piro} {et~al.}(2005){Piro}, {De Pasquale}, {Soffitta}, {Lazzati},
  {Amati}, {Costa}, {Feroci}, {Frontera}, {Guidorzi}, {in't Zand}, {Montanari},
  \& {Nicastro}}]{piro05}
{Piro}, L., {De Pasquale}, M., {Soffitta}, P., {et~al.} 2005, \apj, 623, 314

\bibitem[{{Proga} \& {Zhang}(2006)}]{proga06}
{Proga}, D. \& {Zhang}, B. 2006, \mnras, 370, L61

\bibitem[{{Roming} {et~al.}(2005){Roming}, {Kennedy}, {Mason}, {Nousek}, {Ahr},
  {Bingham}, {Broos}, {Carter}, {Hancock}, {Huckle}, {Hunsberger}, {Kawakami},
  {Killough}, {Koch}, {McLelland}, {Smith}, {Smith}, {Soto}, {Boyd},
  {Breeveld}, {Holland}, {Ivanushkina}, {Pryzby}, {Still}, \& {Stock}}]{uvot05}
{Roming}, P.~W.~A., {Kennedy}, T.~E., {Mason}, K.~O., {et~al.} 2005, \ssr, 120,
  95

\bibitem[{{Rybicki} \& {Lightman}(1979)}]{rybicki}
{Rybicki}, G.~B. \& {Lightman}, A.~P. 1979, {Radiative processes in
  astrophysics}

\bibitem[{{Sari} \& {Esin}(2001)}]{sariesin01}
{Sari}, R. \& {Esin}, A.~A. 2001, \apj, 548, 787

\bibitem[{{Sari} {et~al.}(1998){Sari}, {Piran}, \& {Narayan}}]{sapina98}
{Sari}, R., {Piran}, T., \& {Narayan}, R. 1998, \apjl, 497, L17+

\bibitem[{{Schlegel} {et~al.}(1998){Schlegel}, {Finkbeiner}, \&
  {Davis}}]{schlegel98}
{Schlegel}, D.~J., {Finkbeiner}, D.~P., \& {Davis}, M. 1998, \apj, 500, 525

\bibitem[{{Suwa} \& {Murase}(2009)}]{suwa09}
{Suwa}, Y. \& {Murase}, K. 2009, \prd, 80, 123008

\bibitem[{{Troja} {et~al.}(2007){Troja}, {Cusumano}, {O'Brien}, {Zhang},
  {Sbarufatti}, {Mangano}, {Willingale}, {Chincarini}, {Osborne}, {Marshall},
  {Burrows}, {Campana}, {Gehrels}, {Guidorzi}, {Krimm}, {La Parola}, {Liang},
  {Mineo}, {Moretti}, {Page}, {Romano}, {Tagliaferri}, {Zhang}, {Page}, \&
  {Schady}}]{troja07}
{Troja}, E., {Cusumano}, G., {O'Brien}, P.~T., {et~al.} 2007, \apj, 665, 599

\bibitem[{{Vasileiou}(2013)}]{vlasios13}
{Vasileiou}, V. 2013, Astroparticle Physics, 48, 61

\bibitem[{{Wang} {et~al.}(2006){Wang}, {Li}, \& {M{\'e}sz{\'a}ros}}]{wang06}
{Wang}, X.-Y., {Li}, Z., \& {M{\'e}sz{\'a}ros}, P. 2006, \apjl, 641, L89

\bibitem[{{Yu} \& {Dai}(2009)}]{yu09}
{Yu}, Y.~W. \& {Dai}, Z.~G. 2009, \apj, 692, 133

\bibitem[{{Zhang} {et~al.}(2006){Zhang}, {Fan}, {Dyks}, {Kobayashi},
  {M{\'e}sz{\'a}ros}, {Burrows}, {Nousek}, \& {Gehrels}}]{zhang06}
{Zhang}, B., {Fan}, Y.~Z., {Dyks}, J., {et~al.} 2006, \apj, 642, 354

\bibitem[{{Zhang} {et~al.}(2012){Zhang}, {Burrows}, {Zhang},
  {M{\'e}sz{\'a}ros}, {Wang}, {Stratta}, {D'Elia}, {Frederiks}, {Golenetskii},
  {Cummings}, {Norris}, {Falcone}, {Barthelmy}, \& {Gehrels}}]{zhang12}
{Zhang}, B.-B., {Burrows}, D.~N., {Zhang}, B., {et~al.} 2012, \apj, 748, 132

\end{thebibliography}

\end{document}